\documentclass[twocolumn]{aastex631}
\usepackage{amsmath,amssymb,amstext}
\usepackage{mhchem}
\usepackage{graphicx}
\usepackage{multirow}
\usepackage{booktabs}
\usepackage{float}
\usepackage{appendix}
\graphicspath{{./Figure/}}

\defcitealias{Lustig-Yaeger2023}{Lustig-Yaeger \& Fu et al.}
\defcitealias{Moran2023}{Moran \& Stevenson et al.}
\defcitealias{May2023}{May \& MacDonald et al.}

\submitjournal{ApJ}

\begin{document}

\title{From the Shadows: The Impact of Nightside Thermal Emission on \\ Ultra-hot Jupiter Transmission Spectra Retrievals}

\shorttitle{Nightside Contamination of Hot Jupiter Transmission Spectra}
\shortauthors{Kappelmeier, MacDonald, \& Lewis}

\correspondingauthor{John A. Kappelmeier}
\email{jak485@cornell.edu}

\author[0009-0008-8772-7267]{John A. Kappelmeier}
\affiliation{Department of Astronomy and Carl Sagan Institute, Cornell University, 122 Sciences Drive, Ithaca, NY 14853, USA}

\author[0000-0003-4816-3469]{Ryan J. MacDonald}
\altaffiliation{NHFP Sagan Fellow}
\affiliation{Department of Astronomy, University of Michigan, 1085 S. University Ave., Ann Arbor, MI 48109, USA}
\affiliation{Department of Astronomy and Carl Sagan Institute, Cornell University, 122 Sciences Drive, Ithaca, NY 14853, USA}

\author[0000-0002-8507-1304]{Nikole K. Lewis}
\affiliation{Department of Astronomy and Carl Sagan Institute, Cornell University, 122 Sciences Drive, Ithaca, NY 14853, USA}

\begin{abstract}

Transmission spectroscopy is the most widely used technique for studying exoplanet atmospheres. Since the planetary nightside faces the observer during a transit, highly irradiated giant exoplanets with warm nightsides emit thermal radiation that can contaminate transmission spectra. Observations of ultra-hot Jupiters in the near- and mid-infrared with JWST are especially susceptible to nightside contamination. However, nightside thermal emission is generally not considered in atmospheric retrievals of exoplanet transmission spectra. Here, we quantify the potential biases from neglecting nightside thermal emission in multidimensional atmospheric retrievals of an ultra-hot Jupiter. Using simulated JWST transmission spectra of the ultra-hot Jupiter WASP-33b ($0.8 - 12\,\micron$), we find that transmission spectra retrievals without nightside emission can overestimate molecular abundances by almost an order-of-magnitude and underestimate the dayside temperature by $\gtrsim 400$\,K. We show that a modified retrieval prescription, including both transmitted light and nightside thermal emission, correctly recovers the atmospheric properties and is favored by Bayesian model comparisons. Nightside thermal contamination can be readily implemented in retrieval models via a first-order approximation, and we provide formulae to estimate whether this effect is likely to be significant for a given planet. We recommend that nightside emission should be included as standard practice when interpreting ultra-hot Jupiter transmission spectra with JWST.

\end{abstract}

\section{Introduction}

Transmission spectroscopy \citep{Seager2000,Brown2001} is one of the most frequently used methods to measure the composition and properties of exoplanet atmospheres. Transiting exoplanets obscure a fraction of the host star's light that depends on wavelength due primarily to atmospheric absorption and scattering. Transmission spectroscopy has been successfully used for over two decades to detect atoms and molecules in exoplanet atmospheres \citep[e.g.,][]{Charbonneau2002,Deming2013,Chen2018,Spake2021}. Recently, JWST has delivered high-quality transmission spectra of giant \citep[e.g.,][]{JWSTTransitingExoplanetCommunityEarlyReleaseScienceTeam2023,Grant2023,Radica2023,Xue2023} and rocky exoplanets (e.g., \citetalias{Lustig-Yaeger2023} \citeyear{Lustig-Yaeger2023}; \citealt{Lim2023}; \citetalias{May2023} \citeyear{May2023}) over a longer infrared wavelength range than hitherto possible. The most constraining transmission spectra are generally for hot giant exoplanets, due to their extended scale height producing stronger atmospheric signatures.

Ultra-hot Jupiters ($T_{\rm{eq}} \gtrsim 2000$\,K) are powerful laboratories for atmospheric physics. The strong irradiation experienced by these close-in giant planets drives exceptionally high dayside temperatures, in turn causing a rich diversity of circulation, chemical, and aerosol regimes \citep[e.g.,][]{Fortney2008,Parmentier2018,Helling2019}. Recent years have seen major triumphs from observations of ultra-hot Jupiters, including a plethora of chemical detections \citep[e.g.,][]{Casasayas-Barris2018,Hoeijmakers2018,Tabernero2021,AzevedoSilva2022,Maguire2023,Pelletier2023}, day-night wind speed measurements \citep[e.g.,][]{Ehrenreich2020,Seidel2021,Gandhi2023}, detections of thermal inversions \citep[e.g.,][]{Nugroho2020,Mikal-Evans2022,Yan2022,Coulombe2023}, and temperature maps from both phase curves \citep[e.g.,][]{Kreidberg2018,Arcangeli2019,Mikal-Evans2022} and eclipse mapping \citep{Coulombe2023,Challener2023}. These observational results have been coupled with significant theoretical advances in our understanding of ultra-hot Jupiter atmospheres, including thermal dissociation of molecular hydrogen  \citep{Bell2018,Parmentier2018}, circulation patterns \citep[e.g.,][]{Tan2019,Showman2020} and magnetic fields \citep[e.g.,][]{Hindle2021,Beltz2022,Soriano-Guerrero2023}.

Ultra-hot Jupiter atmospheres are inherently three-dimensional (3D), with large temperature, composition, and cloud variations with altitude, longitude, and latitude. The extreme dayside irradiation experienced by ultra-hot Jupiters drives strong day-night temperature gradients and eastward-flowing equatorial jets \citep[e.g.][]{Showman2002,Menou2009,Rauscher2010,Mayne2014}, alongside substantial chemical gradients, temperature inversions, and asymmetric cloud distributions \citep[e.g.][]{Showman2009,Parmentier2018,Helling2019,Powell2019}. Some of the main sources of chemical gradients in ultra-hot Jupiters include thermal dissociation of molecules such as H$_2$O \citep{Parmentier2018,Lothringer2018} and nightside cold trapping of metal-oxides such as $\ce{TiO}$ and $\ce{VO}$ \citep{Fortney2008,Parmentier2013}. Spitzer observations suggest a nearly universal nightside temperature of $\sim$ 1100\,K for hot Jupiters with $T_{\rm eq} <$ 2500\,K \citep{Beatty2019,Keating2019}, which can be explained by nightside silicate clouds for $T_{\rm eq} <$ 2100\,K \citep{Gao2021}. Higher temperature ultra hot-Jupiters ($T_{\rm eq} \gtrsim$ 2500\,K) are expected to exhibit hotter nightsides due to a relative lack of clouds \citep{Gao2021} and/or the latent heat release from molecular recombination \citep{Bell2018,Roth2021}. Since thermal and chemical gradients can strongly affect transmission spectra \citep{Caldas2019,Pluriel2020,MacDonald2022}, multidimensional models are required to accurately explain ultra-hot Jupiter spectra. 

Multidimensional transmission spectra models have recently been developed for atmospheric retrieval applications. Retrieval codes explore the range of atmospheric properties consistent with an observed spectrum by generating $\gtrsim$ 100,000 model spectra (see \citealt{Madhusudhan2018} and \citealt{Barstow2020a} for reviews). While 3D radiative transfer techniques for transmission spectra have existed for over a decade \citep{Fortney2010,Burrows2010,Dobbs-Dixon2012,Robinson2017a}, the incorporation of multidimensional models into transmission spectra retrievals is a relatively new development \citep{Line2016,MacDonald2017,Lacy2020a,Espinoza2021,Welbanks2021,Welbanks2022,Nixon2022,Zingales2022}. Most recently, \citet{Lacy2020a}, \citet{Nixon2022}, and \citet{Zingales2022} have demonstrated 2D retrieval techniques applicable to ultra-hot Jupiters with strong day-night temperature gradients. However, these existing retrieval techniques have not considered the impact of thermal emission from the nightside of ultra-hot Jupiters.

The nightsides of hot giant exoplanets can contaminate transmission spectra. While an observed transmission spectrum is mainly shaped by transmitted starlight through the planetary terminator, thermal emission from the nightside hemisphere observed before, during, and after a transit can also alter transmission spectra \citep{Kipping2010,Chakrabarty2020,Morello2021}. For exoplanets with cool nightside temperatures, the effect of thermal emission is usually negligible. However, for many hot Jupiters at near and mid-infrared wavelengths --- which can now be probed with JWST --- the assumption of little nightside flux can break down. \citet{Kipping2010} demonstrated that nightside emission can result in a smaller white light transit depth (by a factor $1/(1+F_{\mathrm{p, \, night}}/F_{*})$), reaching 60--80\,ppm at wavelengths $> 5\,\micron$. \citet{Martin-Lagarde2020} later generalized this transit depth bias to consider phase curve variations during transit. \citet{Chakrabarty2020} showed that exoplanets with nightside temperatures exceeding 1200\,K can display significant nightside contamination in transmission spectrum beyond 2\,$\micron$. Most recently, \citet{Morello2021} considered the impact of phase-varying thermal emission on transmission spectra retrievals, finding for simulated JWST transmission spectra of WASP-12b ($T_{\rm{eq}} \approx 2600\,$K) that the H$_2$O abundance retrieved when neglecting thermal emission can be biased by orders of magnitude. However, \citet{Morello2021} assumed 1D transmission spectra with atmospheres in chemical equilibrium. No study to date has investigated the impact of nightside thermal emission on multidimensional atmospheric retrievals.

Here, we explore the impact of nightside thermal emission on multidimensional retrievals of ultra-hot Jupiter transmission spectra. We illustrate this effect via simulated JWST observations of the ultra-hot Jupiter WASP-33b ($T_{\mathrm{eq}} \approx 2800$\,K, $R_{\rm{p}} = 1.50\,R_{\rm{J}}$, $M_{\rm{p}} = 2.1\,M_{\rm{J}}$; \citealt{CollierCameron2010}, \citealt{Lehmann2015}). We selected WASP-33b due to its high estimated nightside temperature ($T_{\mathrm{night}} \approx 1600$\,K; \citealt{vonEssen2020}), its phase curve derived day-night temperature gradient of $\gtrsim$ 1500\,K \citep{Zhang2018,vonEssen2020}, and the brightness of its host star (J = 7.58), leading to a significant transmission spectrum signal-to-noise ratio. Observations of WASP-33b from space-based and ground-based facilities have routinely detected chemical species in its atmosphere via transmission spectroscopy \citep[e.g.,][]{vonEssen2019,Yan2019,Cont2022,vanSluijs2023,Finnerty2023}, rendering it a promising target for nightside emission detectability with JWST. 

Our study focuses on the potential risks and opportunities of nightside thermal emission on transmission spectra retrievals. We first derive the expected signal strength of nightside thermal emission in transmission spectra, before showing the biases from neglecting nightside thermal emission in retrievals of WASP-33b. We also modify the multidimensional transmission model in the POSEIDON\footnote{\href{https://poseidon-retrievals.readthedocs.io/en/latest/}{https://poseidon-retrievals.readthedocs.io/en/latest/}} retrieval code \citep{MacDonald2017a,MacDonald2023} to demonstrate that nightside biases can readily be lifted by including this effect when modelling transmission spectra.

Our study is structured as follows. In Section~\ref{section:derivation}, we derive an analytical expression for the strength of nightside emission on exoplanet transmission spectra. We proceed in Section~\ref{section:wasp33b_model} to calculate a model transmission spectrum of WASP-33b, including nightside contamination, and generate simulated JWST observations. In Section~\ref{section:retrieval_methods} we validate our multidimensional retrieval method for transmission spectra including nightside emission. We demonstrate the impact of neglecting nightside emission on retrievals of simulated JWST data in Section~\ref{section:results}. Finally, in Section~\ref{section:summary_discussion}, we summarize our results and discuss the implications for future near- and mid-infrared JWST observations of hot Jupiters.

\section{The Effect of Nightside Contamination on Exoplanet Transmission Spectra} \label{section:derivation}

\begin{figure*}[hbt!]
    \centering
    \includegraphics[width=0.9\textwidth]{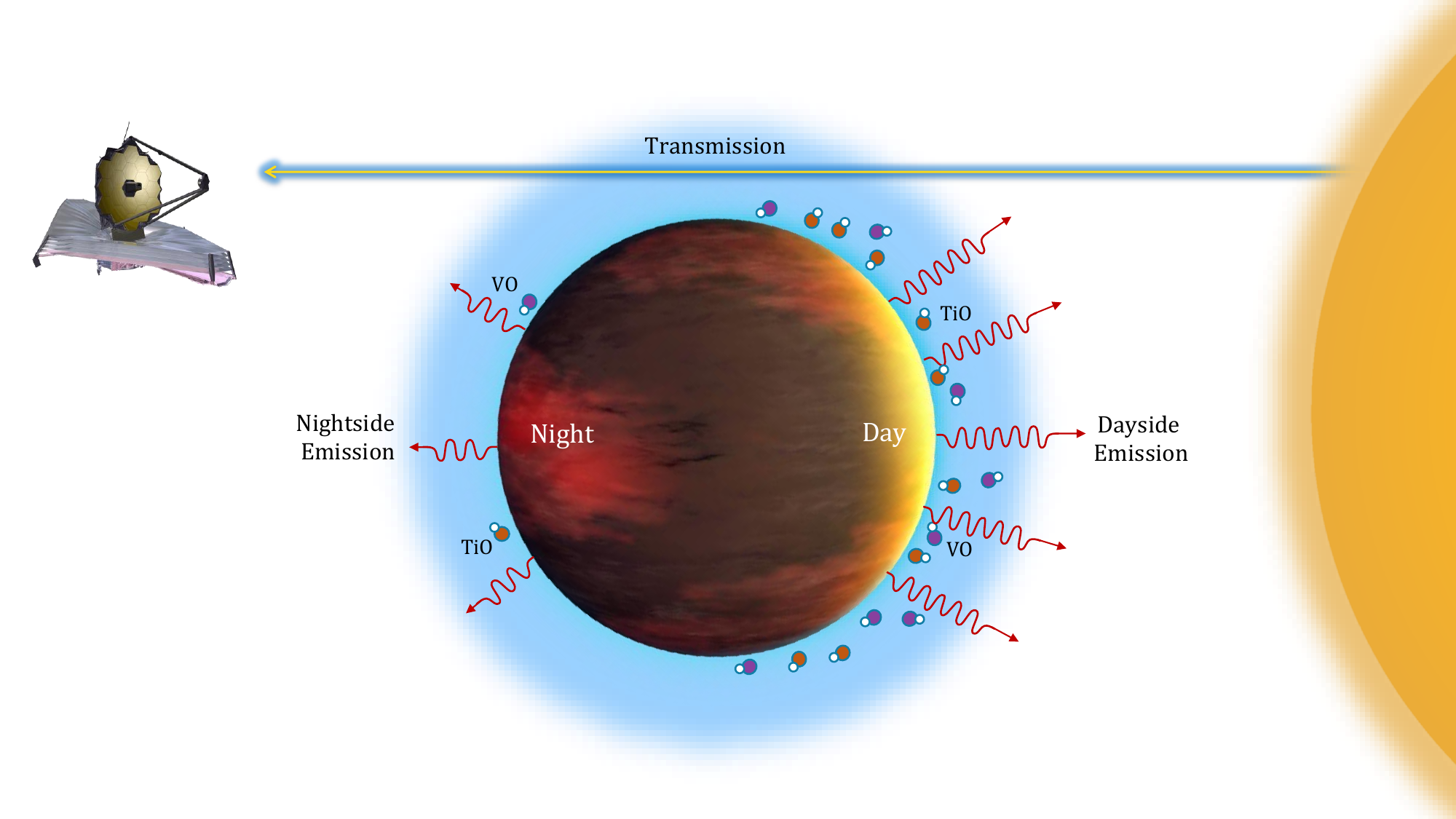}
    \caption{Schematic diagram illustrating the role of nightside emission during an ultra-hot Jupiter transit. The stellar flux from the host star transmits through the exoplanet's atmosphere before reaching the observer, encoding information about its atmospheric properties. The transmission spectrum can be affected by chemical composition gradients between the dayside and nightside, represented here by the relative numbers of $\ce{TiO}$ and $\ce{VO}$ molecules shown on each hemisphere. The observer also receives thermal flux from the planetary nightside. The final transmission spectrum is thus influenced by both the day-night atmospheric property gradients along the path of the stellar rays and the nightside thermal emission. Multidimensional models are therefore required to jointly consider how the nightside affects transmitted starlight and thermal emission. Hot Jupiter concept art adapted from NASA/JPL-Caltech/T. Pyle.
    }
    \label{fig:schematic}
\end{figure*}

We begin by elucidating the mathematical basis for nightside contamination of transmission spectra. A transmission spectrum is primarily composed of contributions from three sources, these being stellar flux transmitted through the planet's day-night terminator, stellar heterogeneities (i.e. star spots and faculae), and thermal emission from the planet's nightside \citep[see][for a derivation and explanation of these terms]{MacDonald2022}. For the sake of this study, we neglect stellar heterogeneities and focus exclusively on nightside emission and transmitted starlight through the terminator. We illustrate these phenomena schematically for a hot Jupiter with a day-night temperature and abundance gradient in Figure \ref{fig:schematic}.

A transmission spectrum including nightside thermal emission can be expressed as follows \citep[e.g.,][]{Kipping2010,MacDonald2022}:

\begin{equation}\label{eq:transmission_factors}
    \Delta_{\lambda} = \frac{F_\mathrm{\lambda,\, out}-F_{\lambda,\, \mathrm{in}}}{F_{\lambda,\, \mathrm{out}}} = \delta_{\lambda,\, \mathrm{atm}}\psi_{\lambda,\, \mathrm{night}}
\end{equation}
where $\Delta_{\lambda}$ is the observed wavelength-dependent transit depth, $\delta_{\lambda,\, \mathrm{atm}}$ is the `standard' transmission spectrum $(R_{\mathrm{p},\, \lambda}^2 / R_{*}^2)$, encoding the wavelength-dependent effective area of the planet relative to the host star, while $\psi_{\lambda, \mathrm{night}}$ is a multiplicative ``contamination factor'' caused by nightside thermal emission.

The first term, $\delta_{\lambda,\, \mathrm{atm}}$, gives the contribution from transmitted starlight to the planet-star effective area ratio. Assuming a planet completely occulting its host star (i.e. not during ingress or egress), for a multidimensional atmosphere this term is given by \citep{MacDonald2022}.

\begin{equation}\label{eq:transmission_atm}
    \delta_{\lambda,\, \mathrm{atm}} = \frac{R_\mathrm{p,\,top}^2 - \frac{1}{\pi}\int_{-\pi}^{\pi}\int_{0}^{R_{\mathrm{p,\,top}}}e^{-\tau_{\lambda}(b, \phi)}\,b\,db\,d\phi}{R_*^2}
\end{equation}
where $R_\mathrm{p,\,top}$ is the radius at the top of the modeled planetary atmosphere, $\tau_{\lambda}$ is the path optical depth, $b$ is the ray impact parameter, $\phi$ is the azimuthal angle, and $R_{*}$ is the stellar radius. Equation~\ref{eq:transmission_atm} solves for the effective area of the planet in a cylindrical coordinate system and holds for 2D or 3D atmospheres \citep[see][for a derivation and further discussion]{MacDonald2022}.

Meanwhile, the multiplicative nightside contamination factor, first derived by \citet{Kipping2010}, is defined as

\begin{equation}\label{eq:nightside_factor}
    \psi_{\lambda,\, \mathrm{night}} = \frac{1}{1 + \displaystyle\frac{F_{\mathrm{p\, (night)}, \, \lambda}}{F_{*, \, \lambda}}}
\end{equation}
where $F_{\mathrm{p\, (night)}\, \lambda}$ and $F_{*,\,\lambda}$ are the observed fluxes of the planetary nightside and star, respectively, as measured at the observer. The observed fluxes are related to the emergent flux from the top of the atmosphere or star (`surface flux') by the scaling relation

\begin{equation}\label{eq:flux_relation_obs_surf}
    F_{\lambda, \,\mathrm{obs}} = \frac{R_{\lambda, \, \mathrm{em}}^2}{d^2} F_{\lambda, \, \mathrm{surf}}
\end{equation}
where $R_{\lambda,\,\mathrm{em}}$ is the radius of the emitting surface ($R_{*}$ for the star and the $\tau \sim 1$ photosphere surface for the planetary nightside) and $d$ is the distance to the observer. Substituting Equation~\ref{eq:flux_relation_obs_surf} into Equation~\ref{eq:nightside_factor}, we obtain the nightside contamination factor in terms of the emergent planet-star flux ratio:

\begin{equation}\label{eq:nightside_factor_2}
    \psi_{\lambda,\, \mathrm{night}} = \frac{1}{1 + \displaystyle\frac{R_{p, \, \mathrm{em}, \, \lambda}^2}{R_{*}^2} \frac{F_{\mathrm{p\, (night), \, surf}, \, \lambda}}{F_{*, \mathrm{surf}, \, \lambda}}}
\end{equation}

We see that the nightside contamination factor, $\psi_{\lambda,\, \mathrm{night}}$, acts to lower the transit depth compared to the usual transmission spectrum formula (Equation~\ref{eq:transmission_atm}). Since the planet-star flux ratio tends to increase with wavelength, the overall effect of $\psi_{\lambda, \, \mathrm{night}}$ is an increasing shift to lower transit depths at longer wavelengths. The factor of $R_{p, \, \mathrm{em}, \, \lambda}^2 / R_{*}^2$ additionally shows that systems with large planet-star radii ratios will maximize the impact of nightside contamination.

To explore what types of planets will have significant nightside contamination, we briefly consider an approximate form of Equation~\ref{eq:nightside_factor_2}. Specifically, we approximate the surface fluxes of the planetary nightside and star as black bodies ($F_{\lambda, \, \mathrm{surf}} \approx \pi B_{\lambda}(T)$), yielding

\begin{equation}\label{eq:nightside_factor_3}
    \psi_{\lambda,\, \mathrm{night}} \approx \frac{1}{1 + \displaystyle\frac{R_{p, \, \mathrm{em}, \, \lambda}^2}{R_{*}^2} \frac{B_{\mathrm{p\, (night)}, \, \lambda}}{B_{*, \, \lambda}}}
\end{equation}
We seek an analytic equation for the bias in the transit depth caused by the omission of nightside thermal emission, which we define as

\begin{align}\label{eq:bias}
    \mathrm{Bias}_\mathrm{night} \equiv \delta_{\lambda,\, \mathrm{atm}} - \Delta_{\lambda} &= \delta_{\lambda,\,\mathrm{atm}}(1 - \psi_{\lambda,\,\mathrm{night}}) \\ \nonumber
    &= \left(\frac{R_{\mathrm{p},\, \lambda}}{R_*}\right)^2 \, \left(1 - \psi_{\lambda,\,\mathrm{night}}\right)
\end{align}
where in the last equality we expressed the regular transmission spectrum, $\delta_{\lambda,\,\mathrm{atm}}$, in terms of the wavelength-dependent effective slant radius of the planet. If we further assume that the emitting radius of the nightside photosphere is approximately equal to both the literature `white light' radius of the planet, $R_p$, and the effective slant radius of the planet ($R_{p, \, \mathrm{em}, \, \lambda} \approx R_{\mathrm{p},\, \lambda} \approx R_p$), we can substitute Equation~\ref{eq:nightside_factor_3} into Equation~\ref{eq:bias} to obtain our analytic approximation for the nightside thermal emission bias

\begin{equation*} \label{eq:nightside_bias_analytic}
    \mathrm{Bias_{night}} \approx \frac{R_\mathrm{p}^2}{R_*^2} \left[ 1 - \left(1 + \frac{R_\mathrm{p}^2}{R_*^2} \displaystyle\frac{ e^{hc / k_B T_* \lambda} -1}{e^{hc / k_B T_{\mathrm{ p\,(night)}} \lambda} -1}\right)^{-1} \right] \tag{8}
\end{equation*}
We see that the effect of nightside thermal emission is largely controlled by the ratio of the squared radii of the exoplanet and its host star, alongside the respective temperatures of the planet and star. It is important to note that Equation \ref{eq:nightside_bias_analytic} only serves as an approximate guide for the estimated bias in the observed transmission spectrum as function of wavelength. Other effects in the atmosphere that are not merely caused by blackbody thermal emission, such as molecular absorption and cloud opacity would produce deviations from this approximation. A more complete assessment of the impact of nightside contamination requires solving the equation of radiative transfer (to calculate the planet-star flux ratio in Equation \ref{eq:nightside_factor_2}) for a specific planetary atmosphere model. This is the focus of the following section.

We now estimate the magnitude of nightside contamination biases across the exoplanet population. For any exoplanetary system with a known planet-star radius ratio, nightside temperature, and stellar temperature, we can calculate the nightside bias using Equation~\ref{eq:nightside_bias_analytic}. To calculate this effect for the entire confirmed transiting exoplanet population ($M_p < 13\,M_\mathrm{J}$), we downloaded the relevant system properties from the exoplanet.eu database. For our purposes, we choose a reference wavelength of $5\,\micron$ (relevant to JWST observations with NIRSpec and MIRI) to estimate nightside thermal emission biases. Since Equation \ref{eq:nightside_bias_analytic} relies on the nightside temperature of the exoplanet, we must estimate the nightside temperature for each of the planets in the database.

\begin{figure}[hbt!]
    \centering
    \includegraphics[trim = 0.2cm -0.5cm 1.0cm -1.0 cm, width=\columnwidth]{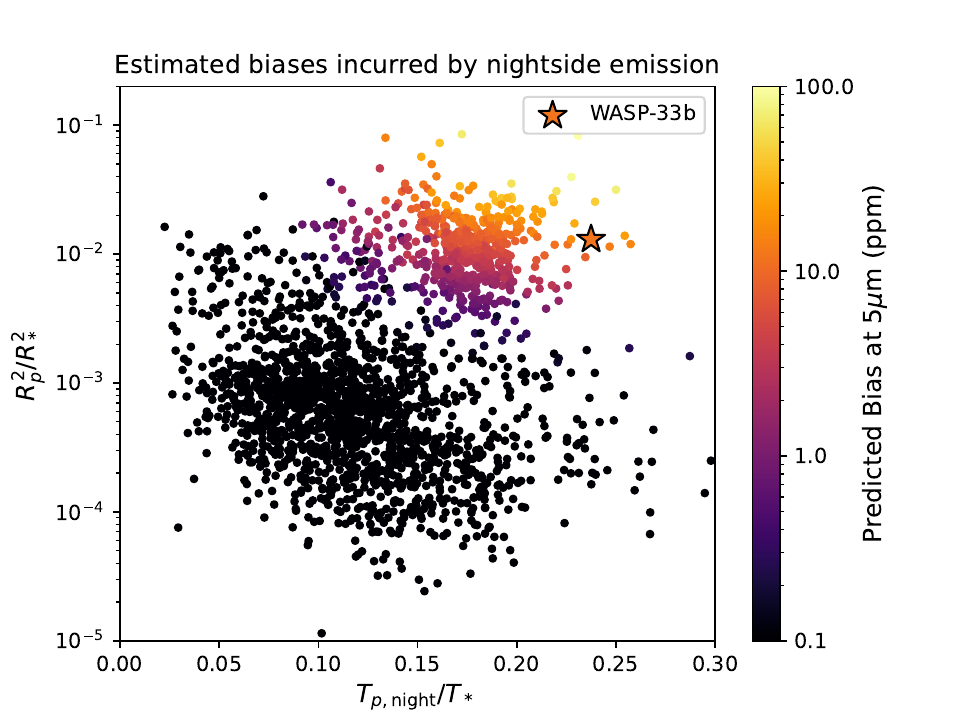}
    \caption{Expected biases in the observed transit depth at $5\,\micron$ for confirmed transiting exoplanets with known planetary radius axis semi-major axis. Each dot represents one of these confirmed exoplanets. The x-axis represents the ratio between the estimated nightside temperature (see Equation \ref{eq:T_p_est}) and the temperature of the host star. The y-axis represents the ratio between the planetary and stellar radius for each exoplanet. Planets with a higher expected bias caused by nightside emission (as calculated from Equation \ref{eq:nightside_bias_analytic}) are shown with a brighter color on a log-scale. The exoplanet we focused our paper on, WASP-33b is indicated on the plot as a star for comparison. Data from Exoplanet.eu (\href{https://exoplanet.eu/}{https://exoplanet.eu/}), accessed on May 14, 2024).}
    \label{fig:biases}
\end{figure}

We construct an approximate scaling relation between the planetary nightside temperature, $T_{\rm p,\,night}$, and equilibrium temperature, $T_{\rm eq}$, using a combination of observational constraints and theoretical insights. First, we note that \emph{Spitzer} observations of many hot Jupiters with $T_\mathrm{eq} < 2100$\,K reveals nightside temperatures nearly uniformly consistent with $\sim 1100$\,K \citep{Beatty2019, Keating2019}. The proposed explanation for this is nightside silicate clouds \citep{Gao2021}. However, for exoplanets with $T_{\rm eq} \gtrsim 2100$\,K, for which silicate clouds shouldn't form \citep{Gao2021}, we assume a linear scaling between $T_{\rm night}$ and $T_{\rm eq}$ with a constant of proportionality calibrated from the observed nightside temperature of the ultra-hot Jupiter WASP-33b ($T_{\rm eq} \sim 2800$\,K, $T_{\rm p,\,night} = 1757$\,K). For colder planets with $T_\mathrm{eq} \lesssim 1100$\,K, where the $\sim 1100$\,K observed nightside temperature constraint \citep{Beatty2019, Keating2019} must break down, we consider the limit of efficient day-night recirculation for a tidally locked planet given by $T_{\mathrm{p,\, night}} = T_\mathrm{eq} \times \pi^{-1/4} \approx 0.75 \, T_\mathrm{eq}$ \citep{Cowan2011}. Since recirculation cannot be more efficient than this relationship, we set a boundary of $T_\mathrm{eq} \geq 1465$\,K for which the silicate cloud temperature of $\sim 1100$\,K holds. Given these insights, we construct a piecewise function to approximate $T_{\rm p,\,night}$.

\begin{equation}
    T_{\mathrm{p},\,\mathrm{night}} = \begin{cases} \label{eq:T_p_est} 
          T_\mathrm{eq} \times \pi^{-1/4} & T_\mathrm{eq}\leq 1465\,\mathrm{K} \\
          1100\,\mathrm{K} & 1465\,\mathrm{K} \leq T_\mathrm{eq} \leq 2100\,\mathrm{K} \\
          T_\mathrm{eq} \times 0.6275 & T_\mathrm{eq} \geq 2100\,\mathrm{K}
       \end{cases} \tag{9}
\end{equation}

Figure~\ref{fig:biases} shows the predicted effect of nightside thermal emission on transmission spectra across the exoplanet population. Generally, systems with high planet-star radii and temperature ratios have the greatest nightside contamination. The size of this effect is $>$ 10\,ppm for many transiting exoplanets, with some predicted to exceed 100\,ppm. For many systems, the large emitting area of the planet relative to the star can dominate over the impact of a higher nightside temperature. Therefore, nightside contamination is not necessarily limited to ultra-hot Jupiters, as $\sim$ Jovian-size planets around small stars can have significant nightside thermal emission.

For the remainder of this study, we select a specific hot Jupiter for a deep investigation of how nightside contamination can impact JWST transmission spectra. We selected the ultra-hot Jupiter WASP-33b for our study, due to it orbiting a bright star with a typical, but not overly optimistic, predicted nightside contamination of 19\,ppm. In what follows, we calculate model transmission spectra and run retrievals for WASP-33b.

\section{WASP-33b: A Case Study in Nightside Contamination} \label{section:wasp33b_model}

Here we consider the ultra-hot Jupiter WASP-33b as a case study of nightside thermal contamination of transmission spectra. In Section~\ref{subsection:forward}, we describe our 2D transmission forward model of WASP-33b, showing cases including and excluding nightside emission. Next, in Section~\ref{subsection:data}, we summarize our JWST data simulations of WASP-33b. These simulated data are used as inputs in our retrieval analysis in the following section.

\subsection{Forward Model} \label{subsection:forward}

\begin{figure*}[hbt!]
    \centering
    \includegraphics[trim = 0.0cm -0.0cm 0.0cm -0.5 cm, width=1.0\textwidth]{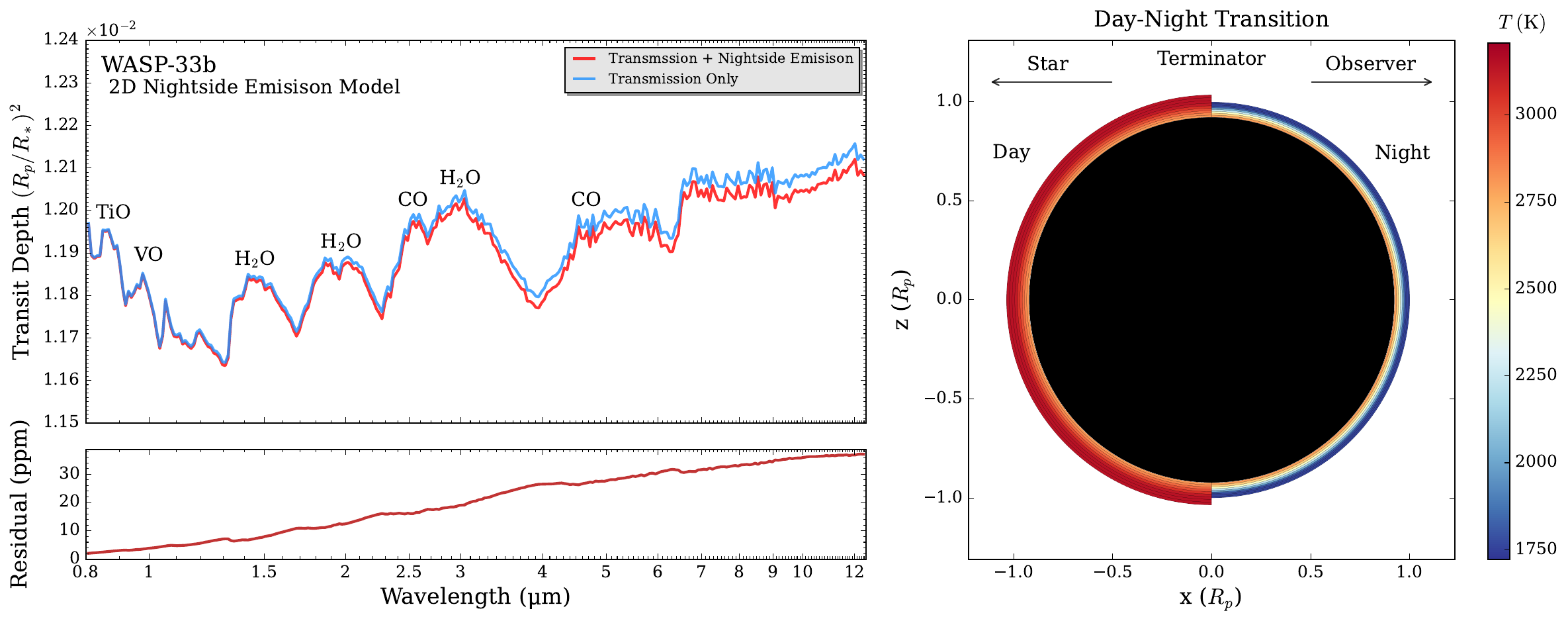}
    \caption{Impact of nightside emission on multidimensional transmission spectra of the ultra-hot Jupiter WASP-33b. Upper left: model transmission spectra of WASP-33b including nightside thermal emission (red) (via Equations~\ref{eq:transmission_factors} and \ref{eq:nightside_factor_2}) and without nightside emission (blue). For clarity, both spectra are shown binned to a spectral resolution of $R = 100$. Lower left: residuals between the transmission spectra with and without nightside thermal emission. The model differences significantly increase with wavelength. Right: 2D geometry of the corresponding WASP-33b model. Stellar rays pass through the day-night terminator, emerging from the nightside, such that the observer sees both transmitted starlight and nightside thermal emission.
    }
    \label{fig:geospec}
\end{figure*}

We created a multidimensional model atmosphere of WASP-33b using the TRIDENT radiative transfer code \citep{MacDonald2022}, which is part of the open source POSEIDON atmospheric retrieval package \citep{MacDonald2017a,MacDonald2023}. We adopt a 2D model for WASP-33b's atmosphere with day-night temperature and composition gradients, assuming azimuthal symmetry around the terminator plane. Our model uses the P-T profile from \citet{MacDonald2022} with $T_{\mathrm{Day}} = 3144\,$K and $T_{\mathrm{Night}} = 1757\,$K \citep[the average Spitzer day and night brightness temperatures of WASP-33b from][]{Zhang2018} for $P < 10^{-5}\,$bar, a common $T_{\mathrm{Deep}} = 2800\,$K for $P > 10\,$bar, and a linear in log pressure temperature gradient connecting $T_{\mathrm{Day}}$ and $T_{\mathrm{Night}}$ to $T_{\mathrm{Deep}}$ in the dayside and nightside, respectively. The dayside and nightside atmospheres span $P = 10^{-9}$--$100\,$bar, with 100 layers spaced uniformly in log-pressure, and an assumed reference radius of at 1.48149\,$R_J$ at $P_{\mathrm{ref}} = 10$\,bar. For simplicity, we assume a sharp transition between the dayside and nightside at the terminator plane. Given WASP-33b's high atmospheric temperature, we assume a cloud-free atmosphere. Figure~\ref{fig:geospec} (right panel) shows the resulting temperature distribution.

For WASP-33b's atmospheric composition, we prescribe a H$_2$-He dominated atmosphere (He/H$_2$ = 0.17) with trace H$_2$O, CO, TiO, and VO. We assume H$_2$O and CO are uniformly distributed throughout the atmosphere, with constant mixing ratios of $\log \rm{H_2 O} = -4.5$ and $\log \rm{CO} = -4$. Our H$_2$O mixing ratio is set lower than the solar value ($\log \rm{H_2 O} \approx -3.3$) as a proxy for thermal dissociation of H$_2$O on the dayside coupled with efficient recirculation (see Appendix~\ref{appendix} for an alternate model explicitly considering a day-night H$_2$O abundance gradient). For TiO and VO, we consider 2D day-night abundance gradients to simulate the expected effect of nightside cold trapping \citep[e.g.][]{Fortney2008,Parmentier2013}. High-resolution observations of WASP-33b have revealed evidence of TiO in its dayside emission spectrum \citep{Nugroho2017,Cont2021} but not at the day-night terminator \citep{Herman2020,Yang2023}, which could potentially offer observational evidence for a day-night TiO gradient. We similarly consider VO due to the recent dayside detection of V \citep{Cont2022}. We adopt dayside abundances of $\log \ce{TiO}_{\mathrm{Day}} = -8$ and $\log \ce{VO}_{\mathrm{Day}} = -8$, while for the nightside we assume a 4 order-of-magnitude depletion ($\log \ce{TiO}_{\mathrm{Night}} = \log \ce{VO}_{\mathrm{Night}} -12$). The adopted atmospheric properties of our reference WASP-33b model are summarized in Table~\ref{tab:properties_and_priors}.

We compute transmission spectra of WASP-33b from 0.8 -- 12.5\,$\micron$ at a resolution of $R = \lambda / \Delta \lambda$ = 10,000. Our radiative transfer employs opacity sampling from pre-computed molecular cross-sections ($\Delta \nu = 0.01$\, cm$^{-1}$; $R \sim 10^6$), where here we use ExoMol line lists for H$_2$O \citep{Polyansky2018}, CO \citep{Li2015}, TiO \citep{McKemmish2019}, and VO \citep{McKemmish2016}. We also include H$_2$-H$_2$ and H$_2$-He collision-induced absorption \citep{Karman2019} and H$_2$ Rayleigh scattering \citep{Hohm1994}. TRIDENT numerically calculates the slant optical depth along the ray path from the dayside to the nightside for 100 rays with impact parameters set to radial distance from the planetary center to the upper layer boundaries on the dayside \citep[see][]{MacDonald2022}. To evaluate the impact of nightside thermal emission, we multiply the `regular' transmission spectrum by $\psi_{\lambda, \, \mathrm{night}}$ (calculated using Equation~\ref{eq:nightside_factor_2}). The emergent planet flux, required to calculate $\psi_{\lambda, \, \mathrm{night}}$, is numerically determined via POSEIDON's thermal emission radiative transfer module \citep[see][their appendix]{Coulombe2023}. For models without nightside contamination, we simply set $\psi_{\lambda, \, \mathrm{night}} = 1$ to reduce to the standard transmission spectrum limit. 

\begin{deluxetable*}{llll} \label{tab:properties_and_priors}
\tablecaption{WASP-33b reference model properties and retrieval priors}
\tablewidth{0pt}
\tablehead{
\colhead{Parameter} & \colhead{Description} & \colhead{Reference Value} & \colhead{Prior Range}
}

\startdata
$R_\mathrm{p,\,ref}$ & Radius at 10\,bar & 1.48\,$R_\mathrm{J}$  & [1.24, 1.70]\\
$T_\mathrm{Day,\,high}$ & Dayside temperature at $10^{-7}$\,bar & 3144 K & [800, 4000]\\
$T_\mathrm{Night,\,high}$ & Nightside temperature at $10^{-7}$\,bar & 1757 K & [800, 4000]\\
$T_\mathrm{Deep}$ & Temperature at 100\,bar & 2800\,K & [800, 4000]\\
$\log \ce{H2O}$ & Water vapor abundance & $-4.5$ & $[-14.0, -0.3]$\\
$\log \ce{CO}$ & Carbon monoxide abundance & $-4.0$ & $[-14.0, -0.3]$\\
$\log \ce{TiO}_{\mathrm{Day}}$ & Titanium monoxide dayside abundance & $-8.0$ & $[-14.0, -0.3]$\\
$\log \ce{TiO}_{\mathrm{Night}}$ & Titanium monoxide nightside abundance & $-12.0$ & $[-14.0, -0.3]$\\
$\log \ce{VO}_{\mathrm{Day}}$ & Vanadium oxide dayside abundance & $-8.0$ & $[-14.0, -0.3]$\\
$\log \ce{VO}_{\mathrm{Night}}$ & Vanadium oxide nightside abundance & $-12.0$ & $[-14.0, -0.3]$\\
\enddata
\end{deluxetable*}

We show the impact of nightside contamination on our WASP-33b transmission spectrum in Figure~\ref{fig:geospec}. Nightside thermal emission decreases the transit depth at longer wavelengths. This intuitive result arises from the planet-star flux ratio increasing with wavelength, such that the nightside emission has a greater dilution effect on the transit at longer wavelengths. We also see this via Equation via Equation~\ref{eq:nightside_factor}, in which an increase in the flux ratio causes $\psi_{\lambda, \, \mathrm{night}}$ to decrease from its non-emission value of 1. For WASP-33b, we see that nightside contamination modifies the transmission spectrum by $\gtrsim$ 30\,ppm at wavelengths beyond 5 $\micron$. Given the typical JWST noise floor of $\lesssim 10$\,ppm \citep[e.g.][]{Schlawin2021,Rustamkulov2022,Lustig-Yaeger2023}, nightside thermal emission could be an important consideration for retrievals of ultra-hot Jupiters, especially for NIRSpec G395H/M or MIRI LRS datasets covering 5$\micron$ and beyond. To explore this possibility quantitatively, we next simulate JWST observations of this model of WASP-33b.

\subsection{Simulated JWST Observations of WASP-33b} \label{subsection:data}

We simulated JWST observations of WASP-33b using the open source PandExo package \citep{Batalha2017a}. We consider the NIRISS SOSS, NIRSpec G395H, and MIRI LRS instrument modes for our purposes. The combination of NIRISS SOSS and NIRSpec G395H is a common choice to maximize the information content for exoplanets transiting bright stars \citet{Batalha2017b}, which consequently improves the efficacy of retrieving atmospheric properties. Due to the high brightness of WASP-33 (J = 7.58), only the first order of NIRSS SOSS can be used in practice. This sets the minimum wavelength of our simulated JWST data at $\approx 0.8\,\micron$. Given that nightside contamination matters more at long wavelengths (e.g. Figure~\ref{fig:geospec}), we also simulated MIRI LRS observations to assess the benefits of MIRI data to constrain nightside emission. For all our PandExo simulations, we assumed a total integration time of 3$\times$ the transit duration (2.85\,hr), a saturation level of 80\% full well, and found the number of groups/integration using PandExo's optimizer. For MIRI LRS, we clipped data points above 12\,$\micron$ due to their low signal-to-noise ratio. We consider 1, 2, and 3 transits with each instrument mode, assuming standard $\frac{1}{\sqrt{N_\mathrm{trans}}}$ error scaling with the number of transits. Finally, we binned the simulated data down to $R = 100$ with the spectres package \citep{Carnall2017} centered on the WASP-33b model spectrum described above (including nightside contamination).  

\begin{figure*}[hbt!]
    \centering
    \includegraphics[width=\textwidth]{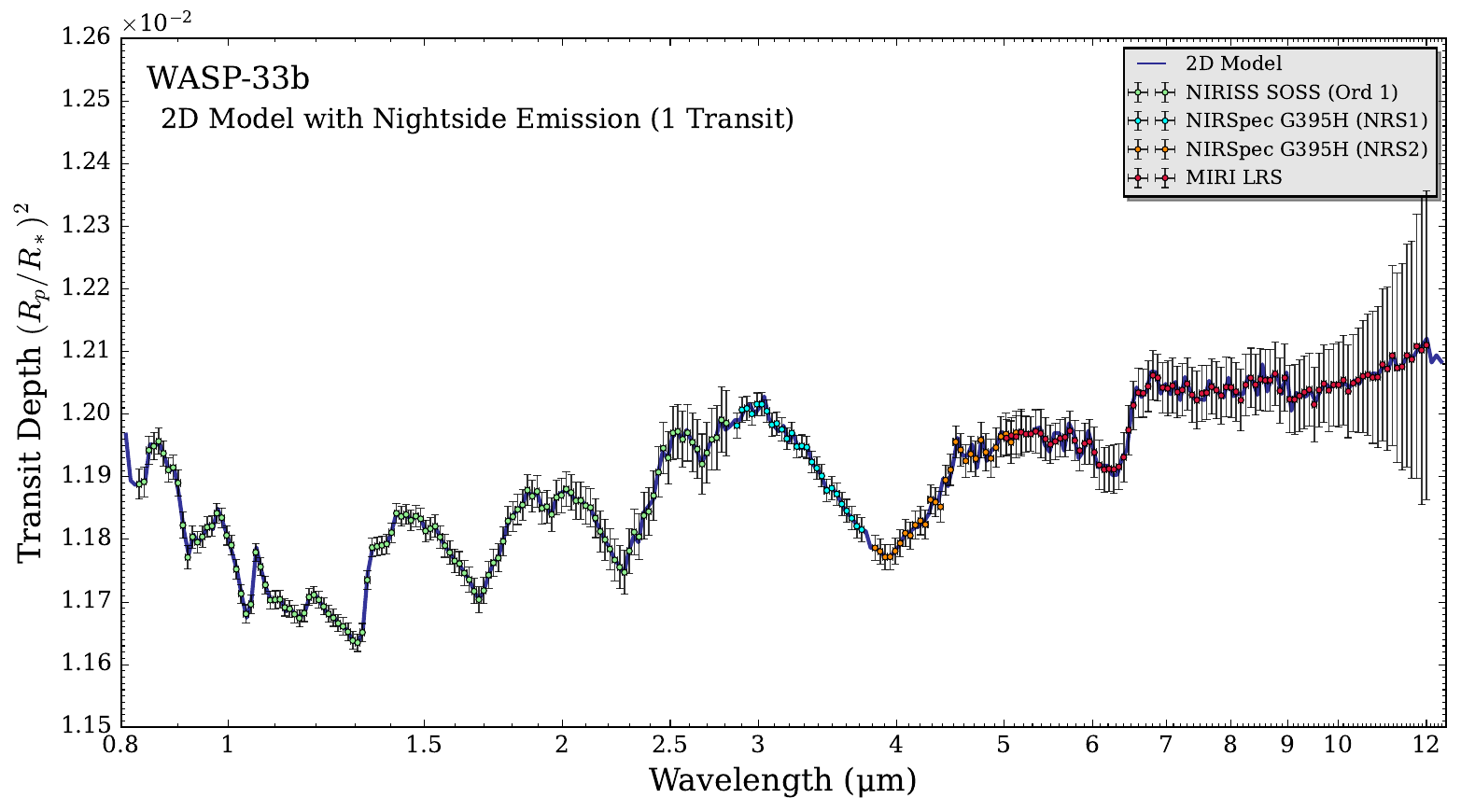}
    \caption{Simulated JWST data for a 2D model transmission spectrum of WASP-33b. Shown are the predicted precisions from PandExo \citep{Batalha2017a}, for a single transit per instrument mode, for NIRISS SOSS (green), NIRSpec G395H (light blue and orange for the NRS1 and NRS2 detectors, respectively), and MIRI LRS (red). The model transmission spectrum corresponds to Figure~\ref{fig:geospec} (dark blue line). All data are binned to $R = 100$ and shown without Gaussian scatter.}
    \label{fig:data}
\end{figure*}

Figure~\ref{fig:data} shows our simulated JWST data of WASP-33b. Each simulated dataset is centred on the predicted observed transit depths of the model (i.e. without Gaussian scatter), after convolution with each instrument's point spread function and integration over the corresponding sensitivity function. We consider data without Gaussian scatter since a retrieval on a dataset without Gaussian scatter is essentially equivalent to posterior averaging over multiple retrievals with different noise instances \citep{Feng2018}. Since our goal is to quantify the average level of biases from nightside contamination on transmission spectra retrievals, this approach mitigates the risk of results specific to a single noise draw.

\section{Transmission Spectra Retrieval Including Nightside Emission} \label{section:retrieval_methods}

We next outline our retrieval approach to include nightside thermal emission when interpreting exoplanet transmission spectra. Our investigation uses a modified version of the open-source multidimensional POSEIDON retrieval framework \citep{MacDonald2017,MacDonald2023}. We describe our retrieval configuration in Section~\ref{subsection:retrieval} before validating that our modified retrieval including nightside emission correctly recovers the true model parameters in Section~\ref{subsection:validation}.

\subsection{Retrieval Configuration} \label{subsection:retrieval}

We consider two different retrieval models for our simulated WASP-33b JWST data. Our first model is a 2D transmission spectra retrieval using the TRIDENT forward model \citep{MacDonald2022} within POSEIDON \citep{MacDonald2017,MacDonald2023}. This retrieval model does not account for thermal emission from the nightside and thus serves as a ``transmission only" model. The second retrieval model alters the forward model to account for nightside thermal emission via Equations~\ref{eq:transmission_factors} and ~\ref{eq:nightside_factor_2}. This is accomplished by calculating a 1D nightside emission spectrum (through integration of the emergent flux from the nightside hemisphere with Gaussian quadrature over the angle $\mu$ between the local normal and the viewing direction), alongside the regular 2D transmission spectrum, for the same 2D atmosphere. Therefore, both the transmission and emission calculations are handled for a common 2D model atmosphere, where the transmitted rays sample both the dayside and nightside while the thermal emission arises solely from the nightside.

Both retrieval models are defined by 10 free parameters, as described in Table~\ref{tab:properties_and_priors}. Our model atmosphere is described by three temperature parameters ($T_\mathrm{Day,\,high}$, $T_\mathrm{Night,\,high}$, and $T_\mathrm{Deep}$) all with uniform priors from 800 to 4000\,K. The mixing ratios of H$_2$O and CO are assumed uniform throughout the atmosphere, with uniform on the logarithm priors from $10^{-14}$ to $10^{-0.3}$. For TiO and VO, we separately parameterize the dayside and nightside abundances with the same priors as the other gases. Finally, we prescribe a uniform prior on the planetary radius at the 10\,bar pressure level from 1.24\,$R_J$ to 1.70\,$R_J$. We assume a sharp transition between the dayside and nightside of the planet for all models for simplicity (the $\beta = 0$ case from \citet{MacDonald2022}). The other model settings, such as the wavelength range and number of layers, are identical to the forward model described in Section~\ref{subsection:forward}. We explore the parameter space using the PyMultiNest \citep{Buchner2014} wrapper for the MultiNest \citep{Feroz2008,Feroz2009,Feroz2019} algorithm.

A notable advantage of using a multidimensional forward model is that no additional free parameters are necessary to describe the nightside emission. Since our 2D atmospheric model already requires a distinct dayside and nightside to calculate transmission spectra, all the necessary atmospheric properties to describe the nightside, and hence to compute emission spectra, are already defined. Without such a multidimensional model, one would have to use separate, uncoupled free parameters to describe the atmospheres at the terminator and nightside. We note that this prescription assumes a common temperature throughout the nightside, such that regions near the anti-stellar point have a similar temperature to the part of the nightside sampled by stellar rays crossing the terminator. Our multidimensional model also has an advantage when conducting Bayesian model comparisons, since the transmission-only and transmission plus nightside emission models have identical priors and parameter space dimensionality. Therefore, any difference in the Bayesian evidence between these models arises solely from including the physics of nightside thermal emission.

Our retrieval investigation considers the impact of nightside contamination for different JWST instrument modes and as a function of observation precision. We first ran 3 retrievals including nightside emission on the full NIRISS SOSS + NIRSpec G395H + MIRI LRS wavelength range (0.8 to 12\,$\micron$), where each retrieval considered 1, 2, and 3 transits per mode (3, 6, and 9 transits total). We ran a further 3 single-instrument retrievals with 1 transit per mode to assess the relative sensitivity of each instrument to nightside contamination. Finally, we repeated each retrieval with a transmission-only model to (i) quantify any biases from omitting nightside contamination and (ii) calculate the Bayesian evidence ratio and hence the detection significance of nightside emission. Our analysis therefore spans a total of 12 retrievals, with all retrievals using 1000 Multinest live points. 

\begin{figure*}[hbt!]
    \centering
    \includegraphics[height=17.4cm]{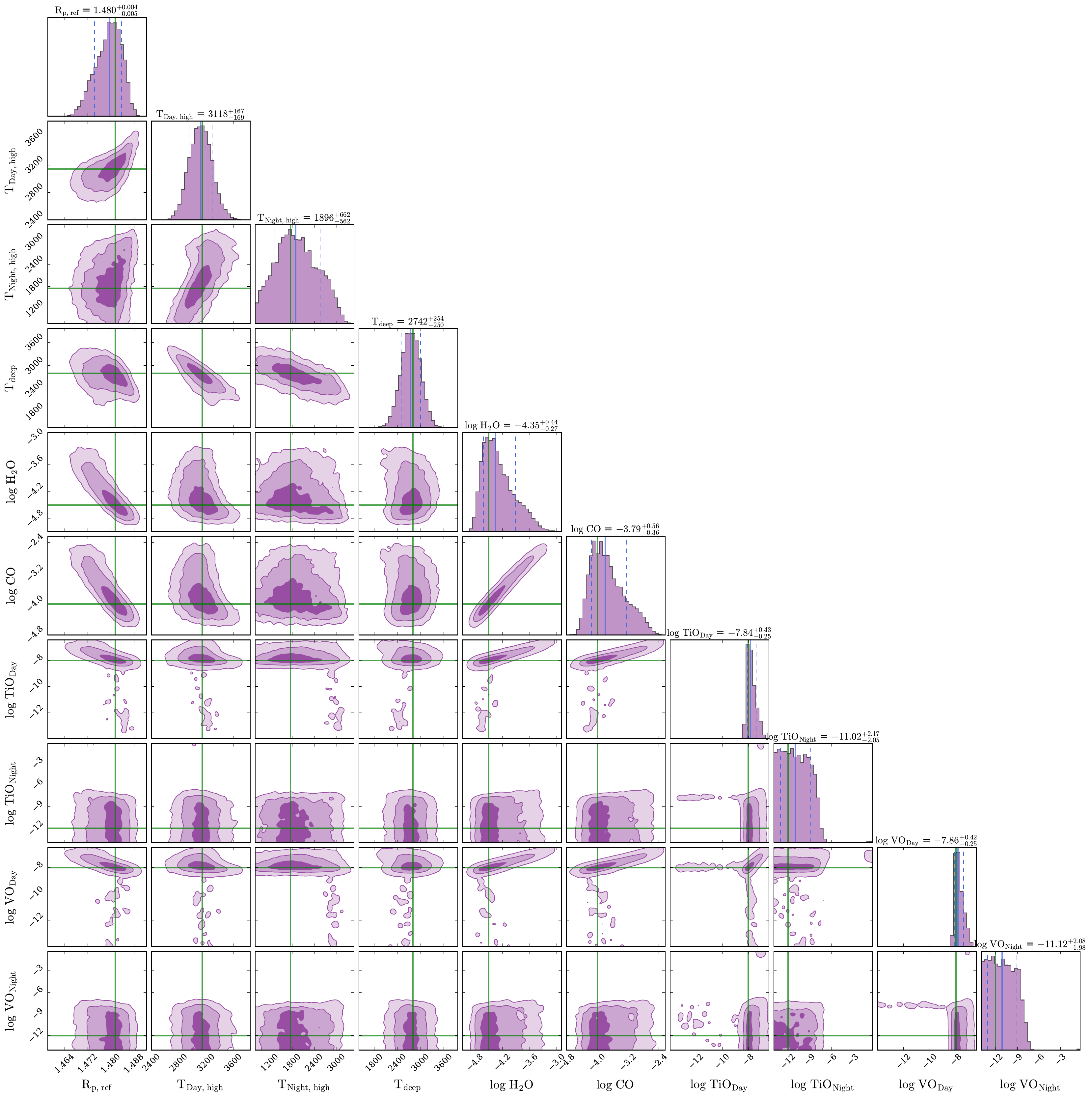}
    \llap{\raisebox{12.0cm}{
      \includegraphics[height=5.4cm]{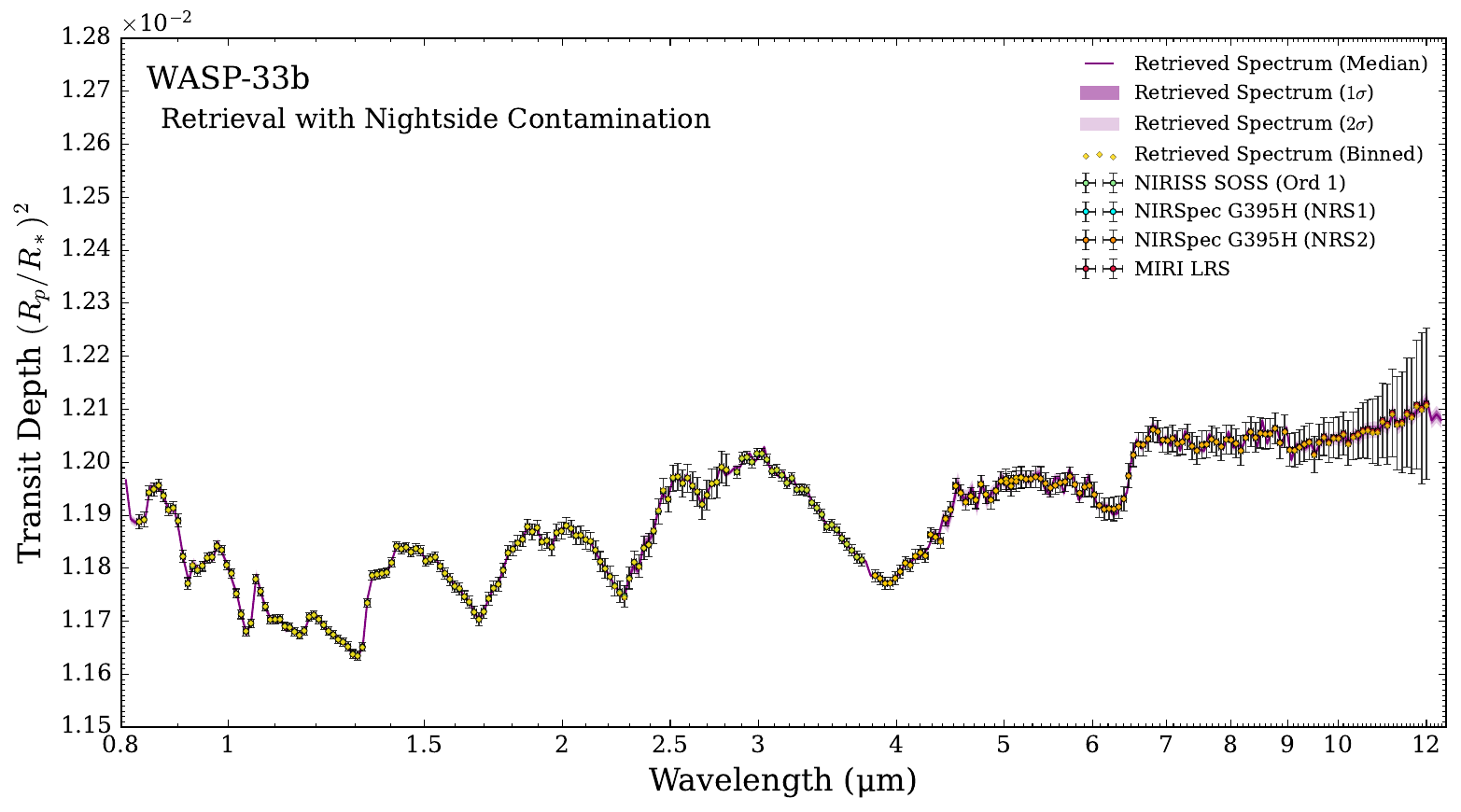}%
    }}
     \caption{2D retrieval validation for a transmission spectrum including nightside thermal emission. Upper right: simulated JWST data for WASP-33b (error bars) compared to the median retrieved spectrum (purple line) and confidence intervals (1 and 2$\sigma$ contours). The model and data are shown at $R = 100$. Lower left: corresponding posterior probability distribution from the retrieval. The purple contours represent the 1$\sigma$, 2$\sigma$, and 3$\sigma$ confidence intervals (in order of increasing opacity). The true parameter values used to generate the simulated JWST data (green lines) compare well with the median retrieved value (solid blue line) and all parameters are correctly retrieved within the 1$\sigma$ confidence interval (dashed blue lines in the histograms). }
    \label{fig:corner_validation}
\end{figure*}

\subsection{Retrieval Model Validation} \label{subsection:validation}

We first validate our modified 2D retrieval framework by comparing the retrieved parameters to the true model inputs. This validation test uses the full simulated data range, including NIRISS SOSS, NIRSpec G395H, and MIRI LRS, with 3 transits per mode. We use this `maximally informative' dataset to validate the retrieval model including nightside contamination, since any issues would show more prominently in the posterior distribution. The true WASP-33b model parameters used in this validation exercise are given in Table~\ref{tab:properties_and_priors}.

Figure~\ref{fig:corner_validation} shows the corner plot and retrieved spectrum from our validation test. We see that all free parameters are successfully retrieved to within 1$\sigma$ of the true value, as expected for retrievals on simulated data without Gaussian scatter \citep{Feng2018}. We also see that the dayside and nightside abundances of TiO and VO are successfully constrained with differing values (bounded constraints for $\log \ce{TiO}_{\mathrm{Day}}$ and $\log \ce{VO}_{\mathrm{Day}}$ but strong upper limits on $\log \ce{TiO}_{\mathrm{Night}}$ and $\log \ce{VO}_{\mathrm{Night}}$ consistent with their lower abundances), demonstrating that day-night chemical composition gradients can be detected and measured from ultra-hot Jupiter transmission spectra with JWST.

We additionally tested the forward model run-time for transmission spectra including nightside contamination. A 2D transmission-only model takes 197\,ms on a single Intel$^{\circledR}$ Core™ i9-11900H CPU @ 2.50\,GHz core to compute a WASP-33b transmission spectrum from 0.8 to 12.5\,$\micron$ at $R$ = 10,000. Our new model including nightside emission takes 459\,ms, driven by the additional computational expense of emission spectrum radiative transfer. This indicates that transmission spectra retrievals including nightside contamination should be roughly 2$\times$ slower than those without. With our retrieval model validated, we proceed to present our results.

\newpage

\section{Results: Retrieval Biases from Nightside Thermal Emission} \label{section:results}

We now demonstrate the impact of neglecting nightside thermal emission from retrievals of transmission spectra. As previously, we use our WASP-33b simulated JWST data (Section~\ref{subsection:data}) for this ultra-hot Jupiter retrieval case study. 
We first consider whether one can infer the missing model complexity of nightside emission from retrieved transmission spectra neglecting this effect (Section~\ref{subsection:results_retrieved_spectra}). We proceed to present how retrieved atmospheric properties change when nightside emission is not included in transmission spectrum retrievals (i.e. retrieval biases; Section~\ref{subsection:results_retrieval_biases}). Our final results investigate the sensitivity of nightside emission inferences to the number of transits (Section~\ref{subsubsection:results_transits}) and the JWST instrument modes included (Section~\ref{subsubsection:results_instruments}). The full posterior distributions for all our retrievals are available online as supplementary material\footnote{\href{https://doi.org/10.5281/zenodo.12629154}{https://doi.org/10.5281/zenodo.12629154}}.

\subsection{Model Complexity: How do we Know Nightside Emission is Required?} \label{subsection:results_retrieved_spectra}

We begin by considering whether a retrieval model not including nightside emission can nevertheless provide an acceptable fit to our WASP-33b simulated JWST dataset including nightside emission. Put another way, can simple visual inspection reveal that nightside emission is missing from a retrieval model? For this purpose, we consider the most optimistic case of 3 transits per observing mode (NIRISS SOSS, NIRSpec G395H, and MIRI LRS) to produce an ultra-precise transmission spectrum from 0.8--12\,$\micron$. 

\begin{figure*}[hbt!]
    \centering
    \includegraphics[width=0.95\textwidth]{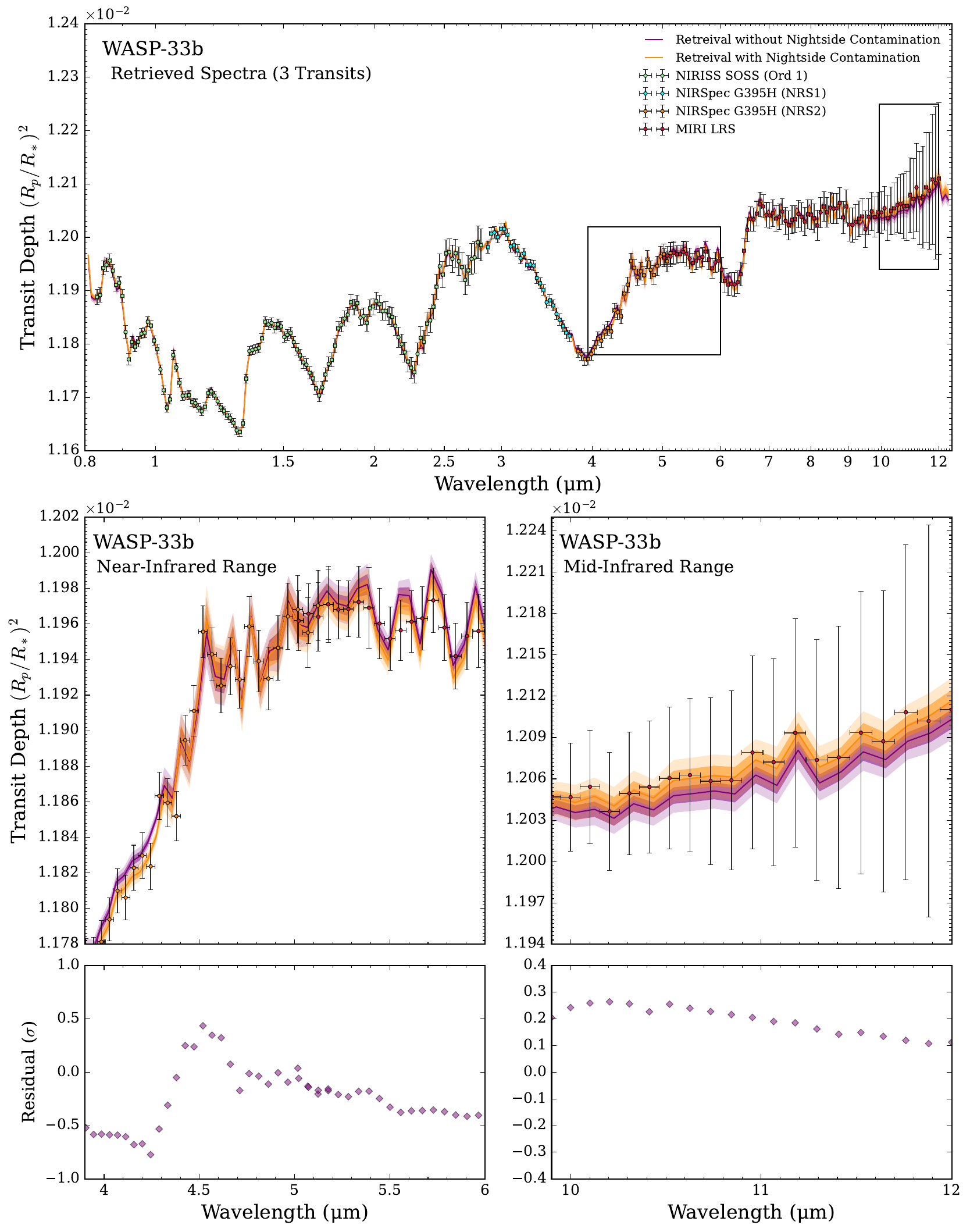}
    \caption{Retrieved transmission spectra for a transmission-only model (purple) vs. a model also incorporating nightside thermal emission (orange). Top panel: the retrieved spectra across the full optical to mid-infrared wavelength range. Middle panels: zoomed-in highlights of two specific regions in the near-infrared ($4\,\micron$ to $6\,\micron$) and mid-infrared ($10\,\micron$ to $12\,\micron$). All model spectra are shown at a resolution of $R = 100$ for clarity. Even though the magnitude of nightside contamination mathematically increases with wavelength (Equation~\ref{eq:nightside_factor}), the transmission-only model can both under-predict (middle left panel) or over-predict (middle right panel) the transit depth compared to the correct model including nightside emission. The bottom 2 panels represent the residuals between the simulated data and the value of the retrieved spectrum not including nightside contamination in units of uncertainty of the data.
    }
    \label{fig:spectracomparison}
\end{figure*}

Figure~\ref{fig:spectracomparison} shows the retrieved transmission spectra for our models including (orange) and not including (purple) nightside emission. We see that the differences between the two retrieved spectra are small --- well within the 1\,$\sigma$ error bars --- such that it is difficult to visually infer the missing physics of nightside emission for WASP-33b. Essentially, the transmission-only retrieval model can compensate somewhat for the missing nightside emission by altering other parameters in the model (producing \emph{biased} retrieval results, as we examine in the next section). However, the lower panels in Figure~\ref{fig:spectracomparison} show that this model compensation is not perfect and does leave small residuals. We show two specific wavelength ranges to illustrate that the transmission-only retrieval can both under-predict the true model transit depth ($4\,\micron$ to $6\,\micron$) or over-predict the transit depth ($10\,\micron$ to $12\,\micron$). In the bottom panels we see that the residuals between the data and the binned model not accounting for nightside contamination are inconsistent with random noise and show significant wavelength correlation. In the $10\,\micron$ to $12\,\micron$ wavelength range, we see a constant underestimation of the retrieved spectrum when not accounting for nightside emission. Since these model differences are too small to enable visual model differentiation, a statistical approach is warranted.

We next consider whether Bayesian model comparisons can differentiate between the transmission-only and nightside emission-included retrieval models. This approach, in essence, integrates over the full wavelength range, combining the information on any mismatch between the model and data, to provide a statistical measure of any preference for one model over the other. For the three transits per mode model comparison, we find a Bayes factor of $3751$ in favor of the model including nightside thermal emission. This Bayes factor is equivalent to a $4.5\sigma$ detection significance for nightside thermal emission vs. the transmission-only model. This demonstrates that Bayesian model comparison offers a powerful avenue to discriminate whether nightside emission is present in a JWST transmission spectrum.

\subsection{Biases: Risks from Omitting Nightside Emission from Transmission Spectra Retrievals} \label{subsection:results_retrieval_biases}

Having shown that transmission-only models can provide a reasonable fit to an ultra-hot Jupiter dataset including nightside emission (see Figure~\ref{fig:spectracomparison}), we next investigate any resulting atmospheric property biases arising from assuming a transmission-only model.

\subsubsection{Retrieved Molecular Abundance Biases} \label{subsubsection:results_abundances}

\begin{figure*}[hbt!]
    \centering
    \includegraphics[width=0.9\textwidth]{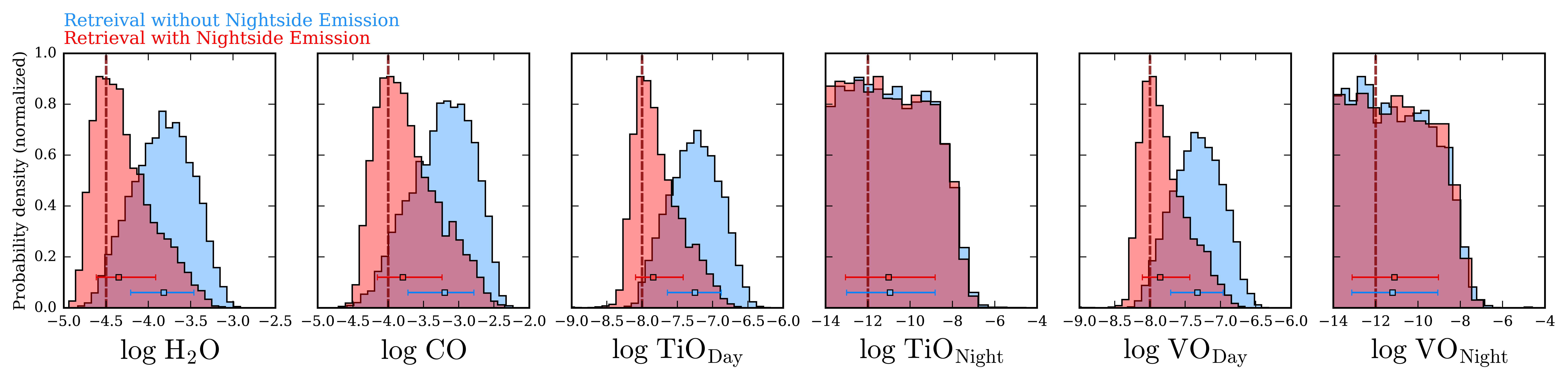}
    \includegraphics[width=0.6\textwidth]{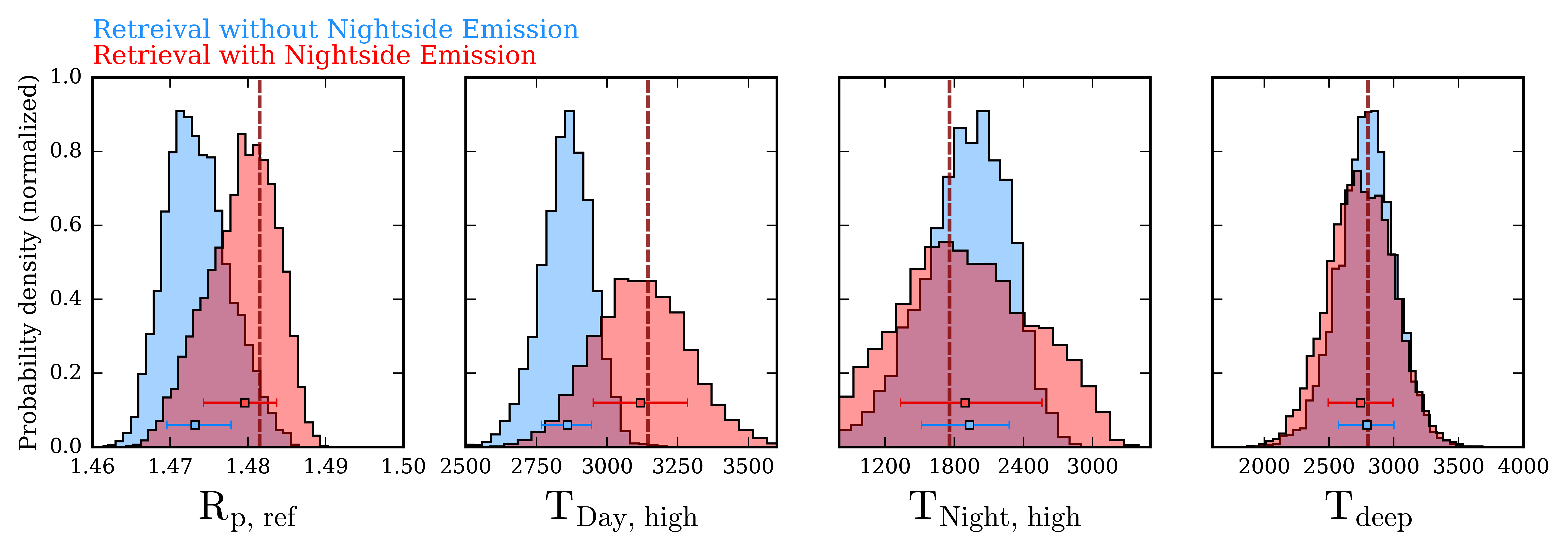}
    \caption{Retrieved atmospheric property biases from neglecting nightside thermal emission. Shown are the posterior distributions from our transmission-only model (blue histograms) and the nightside emission-included model (red histograms) from retrievals of our simulated WASP-33b JWST dataset with 3 transits per mode. Top row: retrieved molecular abundances. Bottom row: retrieved 10-bar reference radius and atmospheric temperature parameters. The true parameter values used to generate the 2D atmospheric model (dashed crimson lines) are overlaid for comparison with the retrieved median and $\pm$ 1\,$\sigma$ value (error bars). The transmission-only model results in notable biases, most notably in the dayside temperature and the abundances for all molecules with bounded constraints (i.e., $\ce{H2O}$, $\ce{CO}$, $\ce{TiO}_{\mathrm{day}}$ and $\ce{VO}_{\mathrm{day}}$).
    }
    \label{fig:biased_histograms}
\end{figure*}

Neglecting nightside thermal emission can incur significant biases in retrieved chemical abundances. Figure~\ref{fig:biased_histograms} illustrates these biases by showing the posterior histograms, for both the transmission-only and thermal emission-included retrieval models, corresponding to our 3 transits per mode simulated WASP-33b dataset. We see clear discrepancies between the true chemical abundances (dashed lines) and the retrieved abundances when nightside emission is neglected (blue histograms), with the abundances of $\mathrm{\ce{H2O}}$ and $\mathrm{\ce{CO}}$, alongside the dayside abundances of $\mathrm{\ce{TiO}}$, and $\mathrm{\ce{VO}}$, biased to higher values. These biases produce discrepancies between the retrieved abundances and their ground truth of  $1$ to $2$\,$\sigma$ for $\ce{H2O}$, $\ce{CO}$, $\ce{TiO}_\mathrm{Day}$ and $\ce{VO}_\mathrm{Day}$, and $>3\,\sigma$ for $\mathrm{T}_{\mathrm{Day},\,\mathrm{high}}$. The largest deviation is for $\mathrm{\ce{CO}}$ ($+0.81$\,dex), while similar magnitude biases are seen for $\mathrm{\ce{H2O}}$ ($+0.69$\,dex), $\mathrm{\ce{TiO}_{Day}}$ ($+0.76$\,dex), and $\mathrm{\ce{VO}_{Day}}$ ($+0.69$\,dex). The retrieved nightside abundances of $\mathrm{\ce{TiO}}$ and $\mathrm{\ce{VO}}$ are not affected by committing nightside emission, as these molecules are not detected on the nightside due to their low nightside abundances.

Including nightside emission in the retrieval model does not notably change the precision of the retrieved parameters. One might intuitively expect that the additional atmospheric information from nightside emission, combined with transmitted starlight, could improve the combined atmospheric constraints. However, Figure~\ref{fig:biased_histograms} shows no significant improvements in the 1\,$\sigma$ precisions for the atmospheric composition parameters. This suggests that any additional atmospheric information contained in the nightside emission spectrum offers little additional constraining power compared to that from the transmission spectrum features. We stress that, although the retrieved composition \emph{precision} may not improve when including nightside emission, this model consideration is still necessary for \emph{accurate} retrieval results.

\subsubsection{Retrieved Pressure-Temperature Structure Biases} \label{subsubsection:results_PT}

\begin{figure}[hbt!]
    \centering
    \includegraphics[width=0.9\columnwidth]{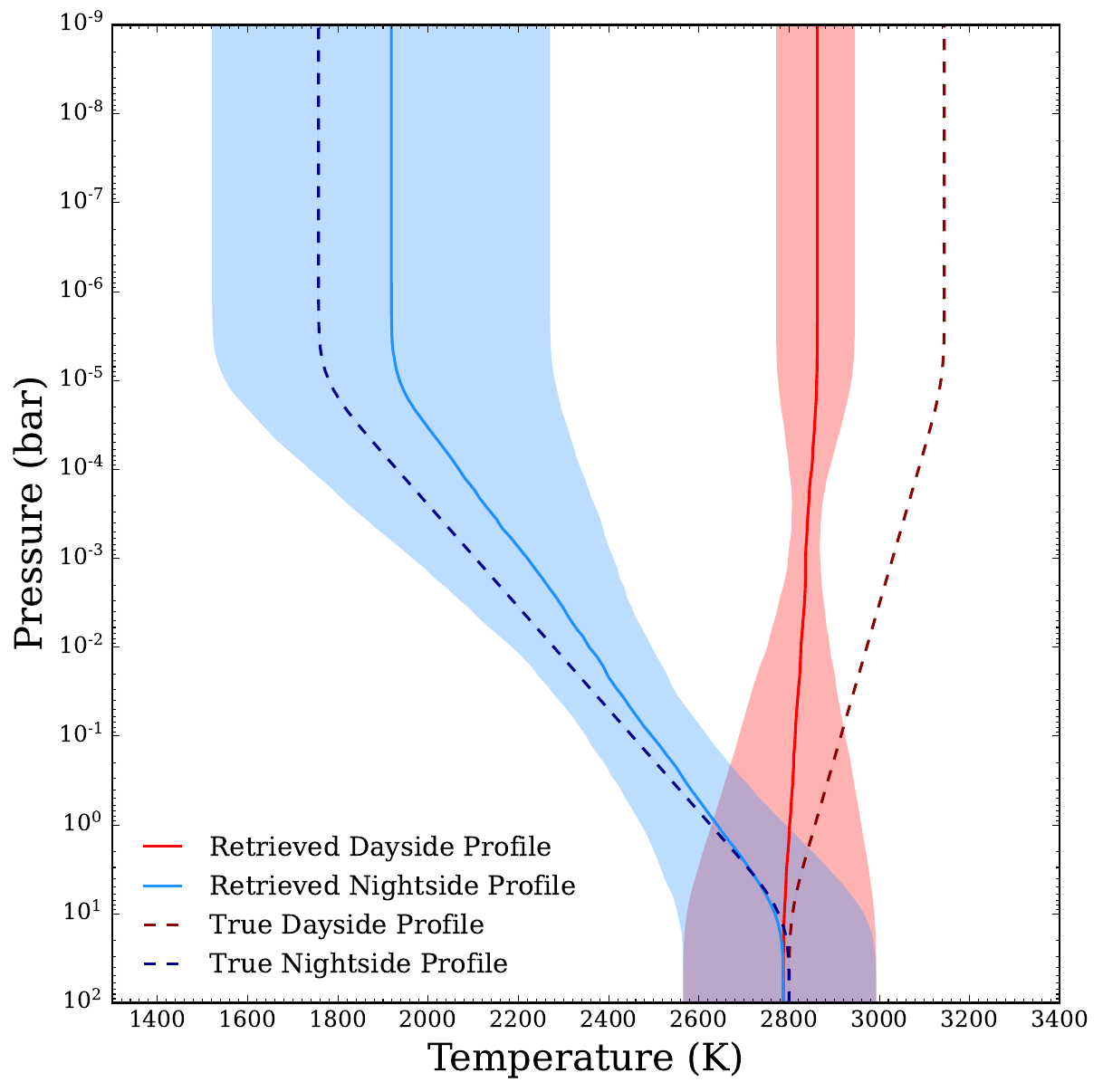}
    \caption{Impact of neglecting nightside emission on retrieved pressure-temperature profiles. The retrieved nightside (blue) and dayside (red) P-T profiles for the transmission-only retrieval model. The true profiles (dashed lines) are compared to the median retrieved profiles (solid lines) and the $1\sigma$ confidence intervals (blue and red shading). These day and night profiles correspond to the blue posterior distributions in Figure~\ref{fig:biased_histograms}. The dayside P-T profile is significantly biased, with a much smaller vertical temperature gradient than the true dayside profile.}
    \label{fig:biased_PT}
\end{figure}
 
Retrieval biases can also affect the inferred temperature structure. We see from Figure~\ref{fig:biased_histograms} that the dayside temperature of WASP-33b is underestimated by nearly $300$\,K when neglecting nightside emission, with the retrieved value $> 2\sigma$ from the ground truth. The retrieved nightside and deep-atmosphere temperatures are relatively unaffected by the omission of nightside emission, being consistent with the true values. We show the retrieved dayside and nightside P-T profiles corresponding to this biased posterior in Figure~\ref{fig:biased_PT}. This plot shows that a key consequence of an underestimated top-of-atmosphere dayside temperature is a correspondingly underestimated day-night temperature gradient.

\begin{figure*}[hbt!]
    \centering
    \includegraphics[width=0.96\textwidth]{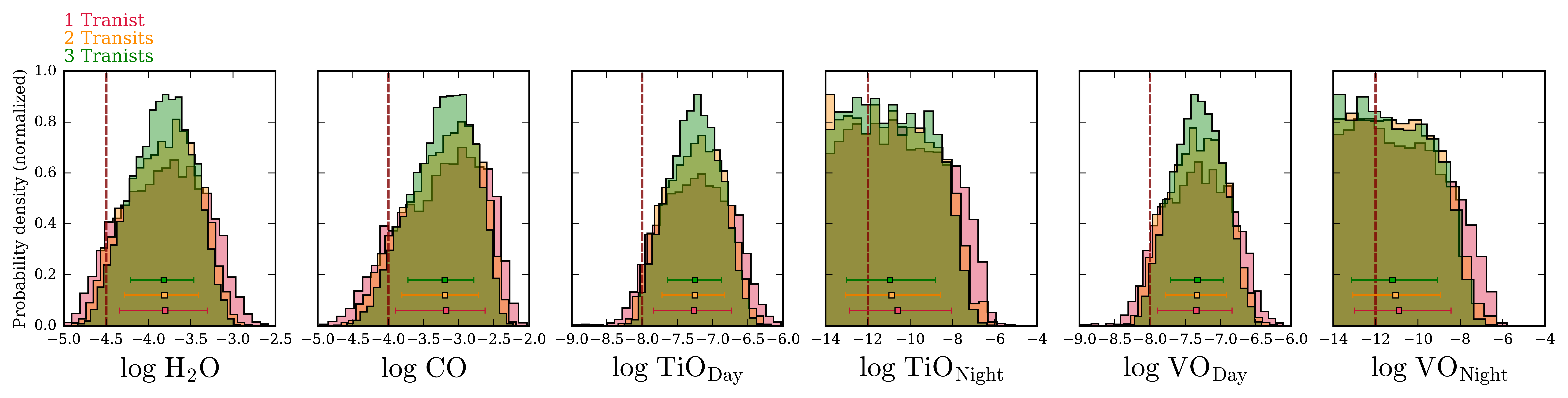}
    \includegraphics[width=0.73\textwidth]{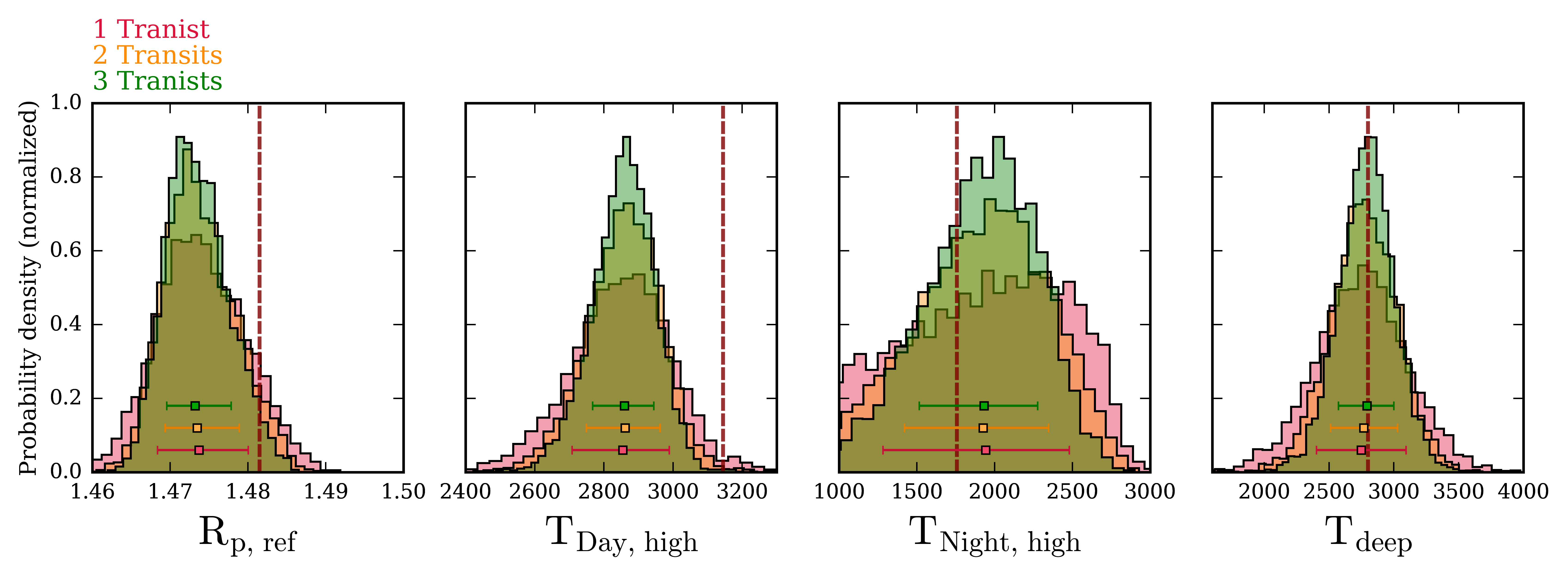}
    \caption{Transit dependence of retrieval biases from neglecting nightside thermal emission from simulated WASP-33b JWST data. The posterior probability distributions for each atmospheric property are shown for 1 transit (red), 2 transits (yellow), and 3 transits (green) per instrument mode (i.e., NIRISS SOSS + NIRSpec G395H + MIRI LRS) compared to the true value of each parameter (crimson dashed line). The median retrieved parameter has almost the same bias regardless of the number of transits, but the biases are more noticeable with additional transits due to the smaller 1\,$\sigma$ confidence intervals.}
    \label{fig:trans}
\end{figure*}

The bias towards lower day-night temperature contrasts is due to the transmission-only model attempting to compensate for the decreasing transit depth at longer wavelengths caused by nightside emission. \citet{MacDonald2022} showed that, for 2D models with day-night temperature and composition gradients, the relative shape of absorption bands can be altered by changing the temperature or abundance gradients. Such an effect could partially mimic the wavelength-dependent transit depth decrease at caused by nightside emission (e.g. Figure~\ref{fig:geospec}). Similarly, the retrieved reference radius for WASP-33b is underestimated by $\approx$ 2\,$\sigma$, which indicates that the transmission-only model is also attempting to slightly decrease the overall transit depth by scaling the amplitude of absorption features. We therefore see that transmission-only retrievals perturb (i.e. bias from the true values) the combination of the chemical composition, day-night temperature gradient, and planetary radius to minimize model residuals caused by nightside emission.  

\begin{figure*}[hbt!]
    \centering
    \includegraphics[width=0.96\textwidth]{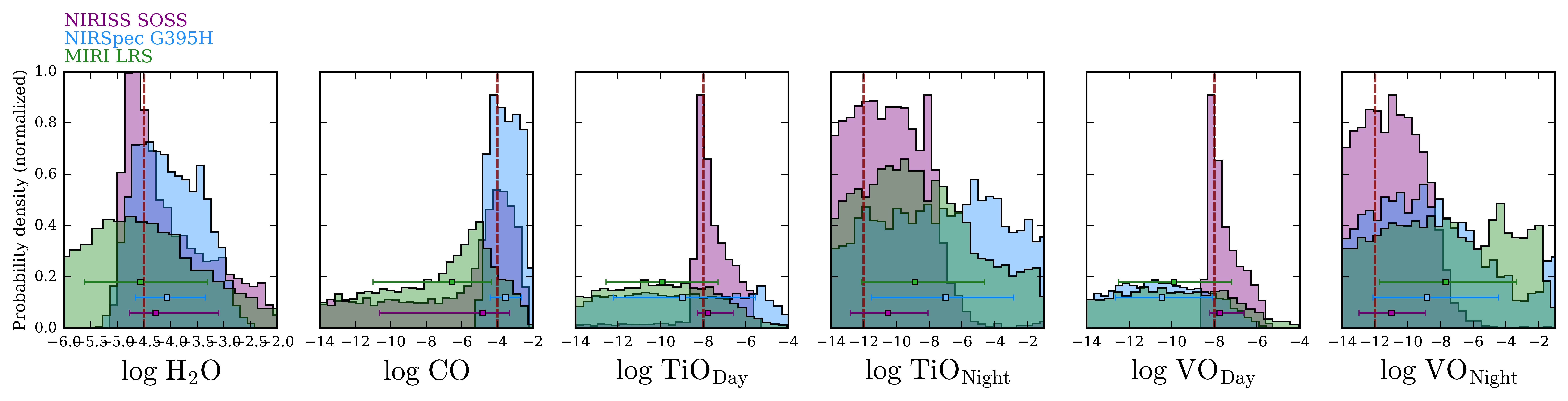}
    \includegraphics[width=0.71\textwidth]{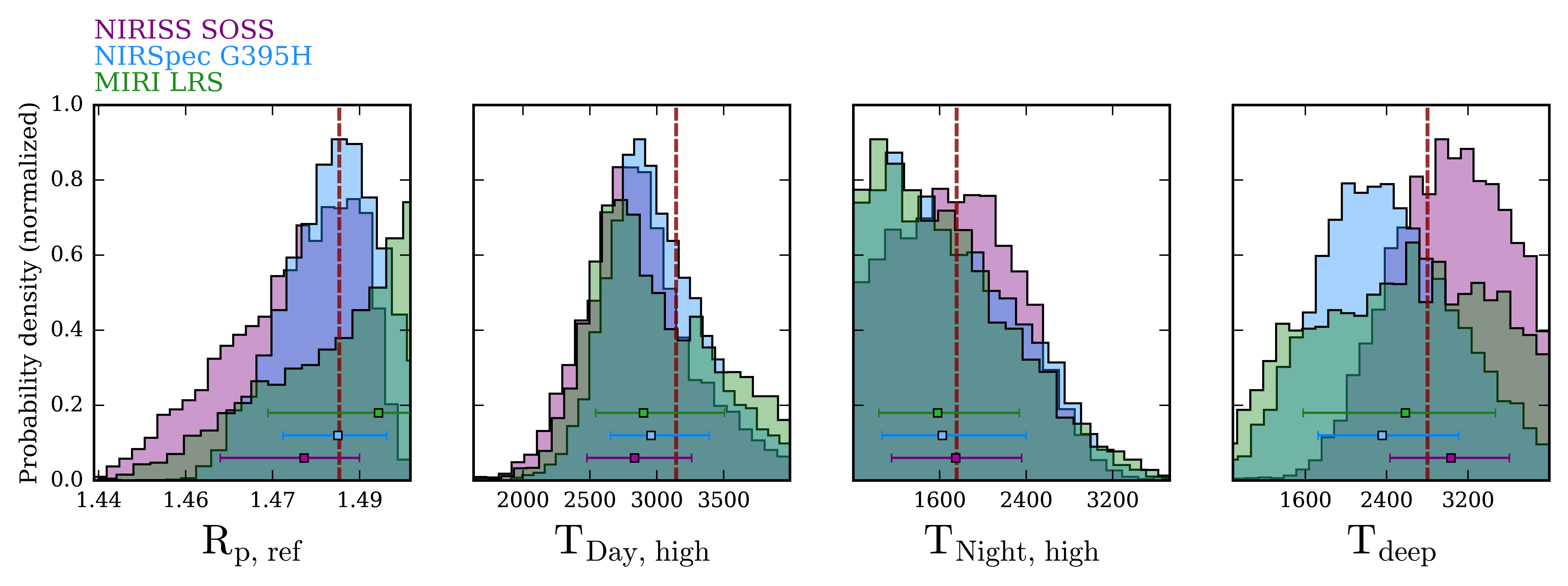}
    \caption{Instrument mode dependence of retrieved atmospheric properties from simulated WASP-33b JWST data. Posterior distributions are shown for 1 transit with a single instrument mode, assuming a transmission-only model, for NIRISS SOSS (purple), NIRSpec G395H (blue), and MIRI LRS (green). The abundances of $\ce{H2O}$, $\ce{TiO}$, and $\ce{VO}$ are best constrained by NIRISS SOSS, while NIRSpec G395H provides the tightest constraints on $\ce{CO}$. Retrieval biases are not seen for single instrument modes, unlike Figure~\ref{fig:biased_histograms}, indicating that a broad wavelength range provides sensitivity to nightside emission contamination.
    }
    \label{fig:instruments}
\end{figure*}

\subsection{Sensitivity to the Number of Transits} \label{subsubsection:results_transits}

Figure~\ref{fig:trans} shows how the biases in our retrieved parameters incurred by nightside emission change with the number of transits per instrument mode. We see that, as expected, an increase in the number of transits for each of the instruments from 1 (red) to 3 (green) corresponds to a narrowing of the posterior distribution. With one transit per mode, most of the constrained atmospheric parameters are biased by $1\,\sigma$ or more - especially the dayside the temperature - with the only relatively unaffected parameters being $\mathrm{T}_{\mathrm{night},\,\mathrm{high}}$, $\mathrm{T}_\mathrm{deep}$, and the nightside abundances of $\ce{TiO}$ and $\ce{VO}$.

The biases incurred from neglecting nightside emission are largely independent of the number of transits observed. Figure~\ref{fig:trans} shows that the abundances of $\ce{H2O}$ and $\ce{CO}$ and the nightside abundances of $\ce{TiO}$ and $\ce{VO}$ are systematically biased by $\approx$ 0.7\,dex higher than their true values. Increasing the number of transits merely makes the impact of the bias more tangible, such that transmission-only retrievals of a more precise transmission spectrum can yield parameter estimates in greater tension with the true underlying atmospheric properties.

We additionally consider how the detection significance for nightside emission varies with the number of transits. For a single transit per instrument mode, we find a Bayes factor of 26 (equivalent to $3.0\sigma$) in favor of the model including nightside emission. Two transits per mode increases the Bayes factor to 176 ($3.7\sigma$), while for three transits per mode the Bayes factor is 3751 ($4.5\sigma$). Our results suggest that a 0.8--12\,$\micron$ JWST transmission spectrum (i.e. NIRISS SOSS + NIRSpec G395H + MIRI LRS) of an ultra-hot Jupiter, with a single transit per instrument mode, offers sufficient information content to moderately favor transmission spectra models including nightside thermal emission.

\subsection{Sensitivity to Instrument Modes} \label{subsubsection:results_instruments}

Figure~\ref{fig:instruments} presents our transmission-only retrieval results for a single transit with only one instrument mode. We see that NIRISS SOSS offers the strongest constraints on the abundances of $\ce{TiO}$, $\ce{VO}$, and $\ce{H2O}$ due to the multiple absorption features of these molecules in the 0.8--2.8\,$\micron$ wavelength range (see Figure~\ref{fig:geospec}). Interestingly, our 2D retrievals with NIRISS SOSS find detections of TiO and VO on the dayside while finding a non-detection on the nightside. This indicates that day-night chemical gradients of TiO and VO are potentially detectable in WASP-33b's atmosphere with even a single NIRISS SOSS transit. In comparison, NIRSpec G395H offers the strongest constraints on CO and an independent (but weaker) constraint on $\ce{H2O}$. Finally, MIRI LRS alone provides only weak constraints on the atmospheric properties. We note that these single-instrument mode retrievals do not contain sufficient information to see biases from nightside contamination, which is hidden within the breadth of their posteriors.

For individual instrument modes, we find it is not possible to detect the presence of nightside thermal emission with a single transit. For all three instrument modes, the Bayes factors are all essentially 1 when comparing retrieval models with and without nightside emission included (indicating no model preference). This contrasts with our finding from Section~\ref{subsubsection:results_transits}) of a 3.0\,$\sigma$ detection of nightside emission for single transits with all three instruments combined. This indicates that data over a wide wavelength range is critical to infer nightside emission from a transmission spectrum. In other words, instruments below $5\,\micron$ provide an important ``anchor'' at wavelengths minimally affected by nightside contamination, enabling retrievals to measure the spectral signatures of nightside emission at MIRI LRS wavelengths.

\section{Summary and Discussion} \label{section:summary_discussion}

In this study, we investigated the effect of nightside thermal emission on our ability to retrieve the atmospheric properties of ultra-hot Jupiters from JWST transmission spectra. We demonstrated, for simulated JWST observations of the ultra-hot Jupiter WASP-33b, that nightside thermal emission can play a sizable role in biasing retrieval results. However, such biases can be readily removed by incorporating nightside thermal emission in a retrieval framework, assuming it is already capable of generating emission spectra and multi-dimensional transmission spectra. This is accomplished by generating two spectra, one for transmission across the day-night boundary and one for thermal emission from the nightside, and combining them via Equations \ref{eq:transmission_factors} and \ref{eq:nightside_factor}. Our key takeaways are as follows:

\begin{enumerate}
    \item \textbf{Biased retrievals:} Exoplanets with large transit depths and high nightside temperatures, most notably ultra-hot Jupiters, can exhibit significant nightside thermal emission. If not accounted for in a retrieval model, this `nightside contamination' can bias retrieved mixing ratios by $\approx 0.7$\,dex.  Nightside emission can also bias retrieved pressure-temperature profiles by artificially decreasing the day-night temperature gradient.
    \item \textbf{Retrievals including nightside emission:} We added nightside thermal emission into the multi-dimensional POSEIDON retrieval framework. Once included in the retrieval forward model, we demonstrated that nightside contamination biases are lifted without a substantial increase in runtime.
    \item \textbf{Detectability of nightside emission:} Separate observations with different instrument modes are required to detect nightside emission from an ultra-hot Jupiter transmission spectrum. It is the \emph{combination} of measurements from instruments both below and above $5\,\micron$, forming a wide wavelength range, that results in a detectable spectral signature of nightside thermal emission. For WASP-33b, one transit with each of our three considered JWST instrument modes (NIRISS SOSS, NIRSpec G395H, and MIRI LRS) would yield a $3.0\,\sigma$ detection of nightside emission. The detection significance rises with additional transits. 
\end{enumerate}

We proceeded to discuss several important implications of our results.

\subsection{Nightside Contamination as a Consideration for Transmission Spectra Retrievals}

Retrievals of ultra-hot Jupiter transmission spectra spanning a wide wavelength range (optical + near-infrared + mid-infrared) should consider nightside emission. Equation~\ref{eq:nightside_factor_3} can be used on a case-to-case basis to provide an initial estimate for the expected magnitude of nightside contamination. Should the expected nightside signature be similar to or greater than the data precision, we strongly advocate for the inclusion of nightside emission in the retrieval forward model. Figure~\ref{fig:biases} demonstrates that the exoplanets with the largest expected bias have large transit depths ($\gtrsim 10^{-2}$) and a planet-star temperature ratio ($\gtrsim 0.15$). We stress that our retrieval results for WASP-33b merely constitute a typical nightside contamination level (25\,ppm), so many planets will exhibit even greater signatures than those considered here. Given the magnitude of the incurred biases seen in our results, and the relative precision of JWST's onboard instruments, retrieval studies for ultra-hot Jupiters not accounting for nightside thermal emission may prove somewhat problematic for accurately determining an atmosphere's molecular composition and day-night temperature gradient. 

Adding nightside emission to a transmission spectrum retrieval requires only a moderate increase in computing resources. Our retrieval model including nightside emission is only $\approx 2\times$ slower than the run-time for a transmission-only retrieval. Though the decision to use a retrieval model accounting for nightside thermal emission will need to be made on a case-by-case basis --- given the properties of the planet and star, the wavelength range of the data, and the data precision --- the computational overhead to account for nightside emission is not overbearing. 

\subsection{Model Limitations and Next Steps}

For the purposes of this study, we made several simplifying assumptions when modeling WASP-33b. First, we limited our study of the multi-dimensionality of WASP-33b to a 2D model with a sharp day-night boundary separating the two hemispheres at the terminator. Consequently, our models assume a uniform temperature (at each altitude) across the nightside hemisphere and a discontinuous temperature change at the terminator. Future retrieval studies could include a parameterized opening angle for the terminator \citep[e.g.][]{Caldas2019,MacDonald2022} or a varying temperature distribution as a function of longitude and latitude across the nightside \citep[e.g.][]{Irwin2020}. Additionally, our 2D model neglected the impact of planetary rotation within the transit \citep[e.g., see][]{Morello2021,Falco2022}.

Second, we restricted our focus to just four trace molecular species, with uniform vertical abundances profiles. In reality, molecular species within hot and ultra-hot Jupiter atmospheres may vary with height. Since the focus of this study was day-night atmospheric gradients, we neglected the effect of altitude-dependent chemical abundances \citep[e.g.][]{Changeat2019}. Additional chemical species, such as H$^{-}$ \citep[e.g.][]{Parmentier2018,Bell2018}, can also provide important sources of opacity in ultra-hot Jupiter spectra.

Third, we assumed a multi-dimensional temperature profile parameterization with a linear in log-pressure temperature gradient. This was used as a first order simplification for a pressure-temperature structure in an exoplanet. However, with the magnitude of nightside thermal emission being a function of the nightside temperature at each altitude, the shape of the temperature profile may impact the results of any multidimensional retrieval. For example, assuming a log-linear temperature profile would not properly capture the temperature profile of an exoplanet with an inversion layer in their atmosphere.

Finally, we neglected the impact of nightside clouds in our retrieval analysis. Since hot Jupiters with equilibrium temperatures above $2100$\,K appear too hot for silicate clouds to form \citep{Wakeford2017, Gao2021}, this is a reasonable approximation for our WASP-33b case study ($T_{\rm{eq}} =$ 2800\,K). Since phase curve observations suggest that clouds proliferate on the nightsides of colder planets \citep[e.g.,][]{Beatty2019,Gao2021}, retrievals assuming clear atmospheres would not be strictly valid for cooler hot Jupiters. However, since nightside clouds effectively act as a surface with a roughly constant emitting temperature \citep{Beatty2019}, we expect that a freely retrieved nightside temperature could nevertheless compensate for the reduced thermal emission from deeper parts of the atmosphere. Future retrieval studies could extend our parameterization to consider the influence of nightside clouds on transmission spectra retrievals of hot Jupiters.

\subsection{Predicted Detectability of Nightside Contamination Across the Exoplanet Population}

\begin{deluxetable}{lcccc}[t!]
\label{tab:estBiases}
\tablecaption{Exoplanets with the largest predicted nightside emission signatures in transmission spectra. }
\tablewidth{0pt}
\tablehead{
\colhead{Planet} & \colhead{$T_\mathrm{eq}$ (K)} & \colhead{J mag} & \colhead{Transit Depth} & \colhead{Nightside} \\[-0.2cm]
\colhead{} & \colhead{} & \colhead{} & \colhead{Precision} & \colhead{Signature} \\[-0.2cm]
\colhead{} & \colhead{} & \colhead{} & \colhead{(ppm)} & \colhead{(ppm)}
}
\startdata 
HIP~65Ab & 1407 & 8.92 & 77.2 & 429.0 \\
WTS-2b & 1533 & 13.46 & 414.0 & 74.3 \\
NGTS-6b & 1525 & 12.22 & 377.1 & 71.7\\
TOI-519b & 752 & 12.85 & 269.3 & 70.7\\
NGTS-10b & 1552 & 12.39 & 287.9 & 66.9\\
Qatar-2b & 1348 & 11.35 & 144.5 & 39.4\\
CoRoT-2b & 1547 & 10.78 & 112.5 & 39.4\\
WASP-43b & 1379 & 10.00 & 92.6 & 39.3\\
TrES-3b & 1643 & 11.02 & 158.3 & 39.1\\
TOI-5205b & 732 & 11.93 & 167.2 & 38.0\\
... & ... & ... & ... & ...\\
WASP-33b & 2781 & 7.58 & 32.2 & 18.8
\enddata
\tablecomments{The transit depth precisions are calculated using PandExo \citep{Batalha2017a}, binned to a typical spectral resolution of $R = 100$, and quoted at 5\,$\micron$. The 10 exoplanets are ordered according to their predicted nightside impact on transmission spectra at $5\,\micron$ (estimated via Equations~\ref{eq:nightside_bias_analytic} and \ref{eq:T_p_est} -- see Section~\ref{section:derivation}). We include WASP-33b, the case study planet investigated in this study, for comparison. Planet properties are from exo.MAST (\href{https://exo.mast.stsci.edu/}{https://exo.mast.stsci.edu/}).}
\end{deluxetable}

Our study focused solely on the case study of the ultra-hot Jupiter WASP-33b. This served as a representative example among exoplanets where nightside thermal emission is predicted to play a relatively large role (see Figure~\ref{fig:biases} and Table~\ref{tab:estBiases}). Here, we briefly consider other planets where nightside thermal emission could be detectable. The detectability of nightside emission is a balance between the intrinsic strength of the signal (i.e. Equation~\ref{eq:nightside_factor_3}) and the brightness of the host star).

First, we calculated a ranked list of the top 10 exoplanets with the greatest predicted nightside emission signature, shown in Table~\ref{tab:estBiases}. We see that nearly all of the top 10 targets are large gas giants with equilibrium temperatures over 1000\,K. However, the top targets are not limited solely to ultra-hot Jupiters, with many typical temperature ($T_{\rm{eq}} \approx$ 1500\,K) hot Jupiters among the top targets --- since the absolute transit depth (i.e. the planet-star relative emitting area) matters more than the nightside temperature. We highlight specifically that the ultra-short-period hot Jupiter HIP~65Ab \citep{Nielsen2020} is predicted to have a substantial nightside emission signature of over 400\,ppm at 5\,$\micron$ ($>$ 16\,$\times$ stronger than for WASP-33b).

Second, we used PandExo \citep{Batalha2017a} to compare these predicted nightside emission signatures with the typical transit depth precision for each of these exoplanets. We simulated single transit JWST observations with the commonly used NIRSpec G395H instrument mode, chosen due to its wavelength range encompassing where nightside emission is prominent (near and beyond 5\,$\micron$) and its high instrument throughput. The J band magnitudes for each of the exoplanets' host stars in Table~\ref{tab:estBiases} were taken from the SIMBAD database \citep{Wenger2000}. Our PandExo simulations all assumed a total integration time of 3$\times$ the transit duration, a saturation level of 80\% full well, and the default number of groups/integration from PandExo's optimizer. We binned the simulated NIRSpec G395H observations down to $R = 100$ and quote the transit depth precision for the data point nearest 5\,$\micron$ in Table~\ref{tab:estBiases}.

As we can see from Table~\ref{tab:estBiases}, there are many exoplanets that have predicted nightside contamination biases that are larger than WASP-33b. However, many of these exoplanets are located around relatively dim stars, resulting in a significantly worse nightside emission signal-to-noise ratio compared to WASP-33b. However, we see that the planet with the largest predicted nightside signature, HIP 65Ab, also has a host star with comparable brightness to WASP-33. Consequently, the expected nightside emission signal-to-noise ratio for HIP 65Ab is expected to be nearly 10$\times$ that WASP-33b. We therefore suggest that HIP 65Ab would make an excellent target to confidently detect nightside emission from a hot Jupiter with JWST. Such observations would also prove critical to further refine multi-dimensional retrieval techniques, such as the method demonstrated in this study.

We have shown that nightside thermal emission can be important in sculpting exoplanet transmission spectra and atmospheric retrieval results. However, nightside contamination need not prove a ghastly spectre harming our interpretation of JWST transmission spectra. By including the emergent flux from the shadowy nightside in retrievals, we can rest assured that we infer an accurate picture of the atmospheres of ultra-hot Jupiters.

\section*{Acknowledgments}

R.J.M. is supported by NASA through the NASA Hubble Fellowship grant HST-HF2-51513.001, awarded by the Space Telescope Science Institute, which is operated by the Association of Universities for Research in Astronomy, Inc., for NASA, under contract NAS 5-26555. We thank the anonymous referee for helpful feedback that improved the quality of our study.

\newpage

\appendix

\section{Impact of H$_2$O Dissociation} \label{appendix}

Here we examine an altered simulated model of WASP-33b's atmosphere. In Section \ref{section:wasp33b_model} we modeled WASP-33b as an exoplanet whose $\ce{H2O}$ abundance was constant throughout the atmosphere, unlike $\ce{TiO}$ and $\ce{VO}$ which was modeled with a day-night abundance gradient. Since many ultra-hot Jupiters often exhibit differences in their $\ce{H2O}$ abundance between their dayside and nightside due to $\ce{H2O}$ destruction on the dayside and recombination on the nightside (source). All the parameters of the atmosphere from Table \ref{tab:properties_and_priors} were kept the same, except for $\ce{H2O}$ which was now modeled with a log abundance of $-4.0$ and $-2.0$ on the dayside and nightside respectively.

\begin{figure*}[hbt!]
    \centering
    \includegraphics[scale=.6]{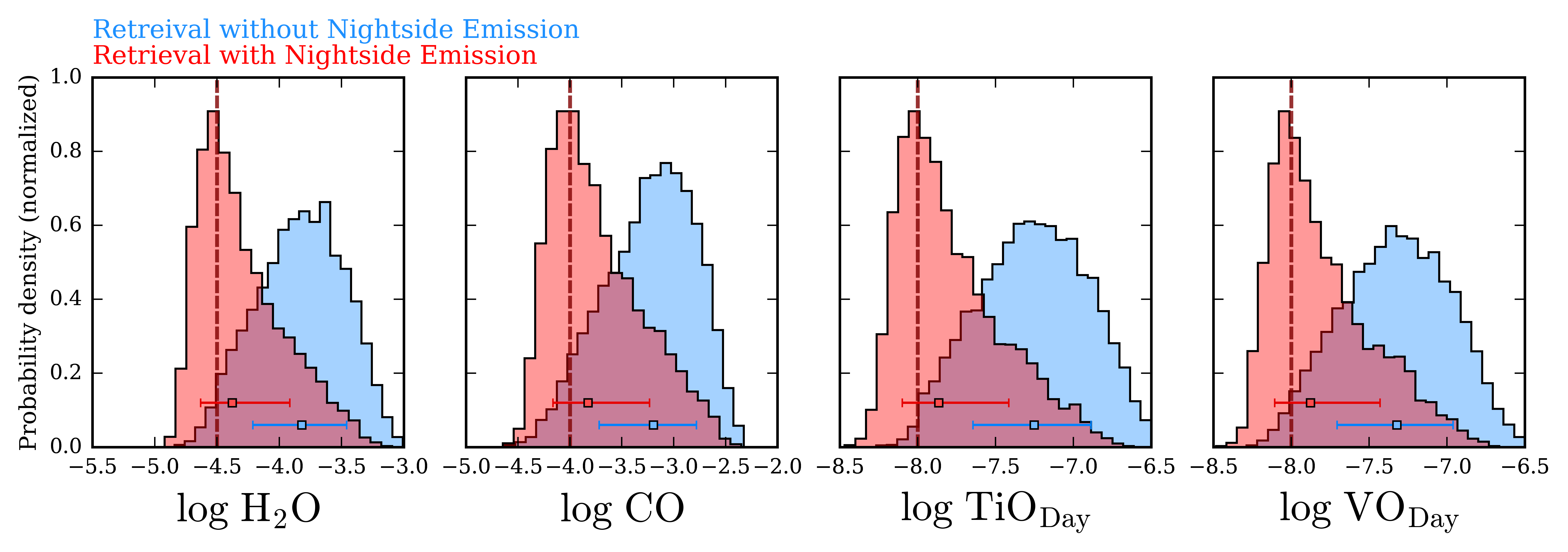}
    \includegraphics[scale=.6]{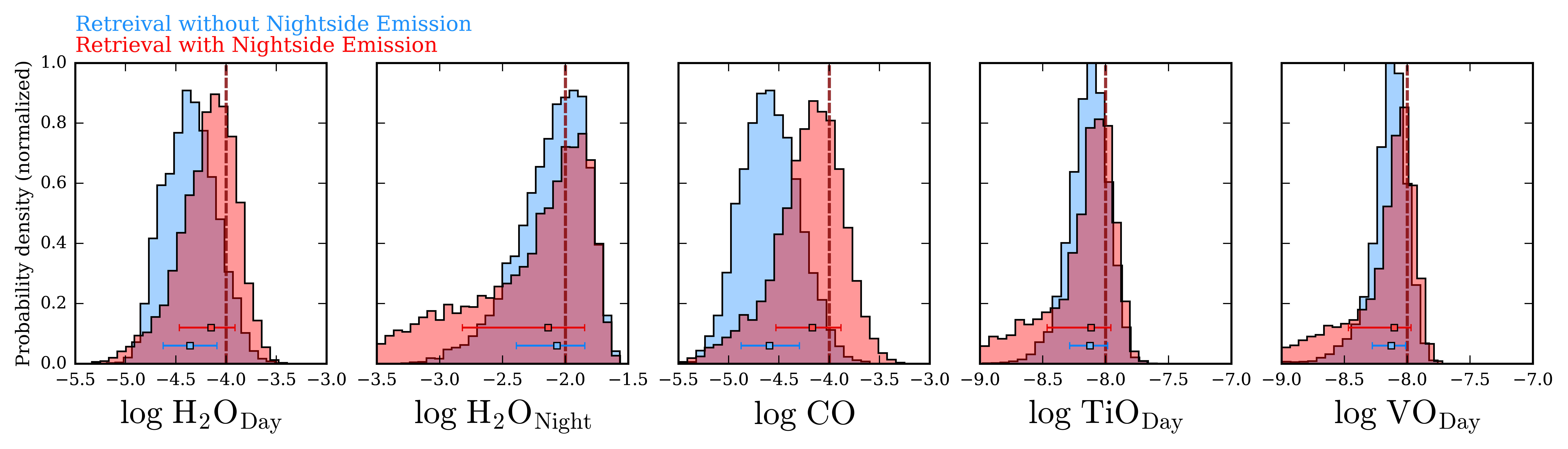}
    \caption{Retrieved posterior distributions for various atmospheric properties of WASP-33b for a simulated 3 transit observation. Top: retrieved posterior probability distribution for a model of WASP-33b with a 1-dimensional distribution of $\ce{H2O}$. This is equivalent to the distributions seen in Figure \ref{fig:biased_histograms}. Bottom: retrieved posterior distributions for a model of WASP-33b with a difference in the abundance of $\ce{H2O}$ between the dayside and nightside. The blue distributions indicate the results for a retrieval that does not account for nightside thermal emission, while the red shows the results for the retrieval models including nightside emission. The dashed crimson lines indicates the true value of each parameter which was used to generate the simulated spectrum. }
    \label{fig:comparison}
\end{figure*}

The comparison of the results of these two atmospheric scenarios of WASP-33b are shown in Figure \ref{fig:comparison}. As we can see, unlike for the 1-dimensional $\ce{H2O}$ scenario, it is evident that the effect that nightside thermal emission has on biasing the retrieval results does not always work in one direction. The scenario with constant $\ce{H2O}$ results in the retrieval overestimating the abundance by nearly 1 order of magnitude. However, in the day-night $\ce{H2O}$ transition scenario, the opposite is true, and instead nightside emission works to result in an underestimation of the abundance of the molecule on both sides of the planet. This is similarly true with the estimated abundance of $\ce{CO}$ in both portions of the figure. This small change in the atmospheric model demonstrates that while in our main scenario, the abundances for all species were overestimated by not accounting thermal emission in the retrieval model, this was just a quirk of the scenario and it can work in either direction. 

\bibliography{nightside_contam}
\bibliographystyle{aasjournal}

\end{document}